\documentclass[11pt,letterpaper]{article}
\ifdefined\pdfminorversion\pdfminorversion=7\fi

\usepackage[margin=1in]{geometry}
\usepackage[T1]{fontenc}
\usepackage{amsmath,amsthm}
\usepackage{newtxtext}
\usepackage{newtxmath}
\usepackage{microtype}
\usepackage{array,graphicx,placeins}
\usepackage[font=small,labelfont=bf]{caption}
\usepackage{tikz,pgfplots}
\usetikzlibrary{arrows.meta,positioning}
\usepgfplotslibrary{groupplots}
\pgfplotsset{compat=1.18}
\usepackage[authoryear,round]{natbib}
\usepackage{chapterbib}
\usepackage[hidelinks]{hyperref}

\graphicspath{{figures/}}
\DeclareMathOperator{\pr}{pr}

\DeclareMathOperator{\Pa}{Pa}
\def\T{{\mathrm{\scriptscriptstyle T}}}

\newtheorem{theorem}{Theorem}
\newtheorem{lemma}{Lemma}
\newtheorem{corollary}{Corollary}
\newtheorem{proposition}{Proposition}
\theoremstyle{definition}
\newtheorem{assumption}{Assumption}

\newenvironment{keywords}
  {\par\smallskip\noindent\small\textit{Some key words}: \ignorespaces}
  {\par\medskip}
\newcommand{\tbl}[2]{\caption{#1}\centering{\small #2\par}}
\newenvironment{tabnote}
  {\par\smallskip\small\raggedright\noindent\ignorespaces}
  {\par}

\newcommand{\PaperTitle}{Sharp bounds for posterior causal attribution given complete factual records}
\newcommand{\PaperAuthors}{Liyao Zhou and Xinwei Sun}

\title{Sharp bounds for posterior causal attribution\\
given complete factual records}
\author{\PaperAuthors\\[0.35em]
  \small\href{mailto:liyaozhou04@gmail.com}{liyaozhou04@gmail.com}}
\date{}
\hypersetup{pdftitle={\PaperTitle},pdfauthor={\PaperAuthors}}

\begin{document}
\raggedbottom
\maketitle
\begin{abstract}
Retrospective causal attribution evaluates whether altering an earlier event would have prevented an observed outcome. Observational data determine marginal potential-response probabilities at fixed inputs while leaving cross-world joint distributions unspecified. Conditional on a complete factual record in a finite Markovian structural causal model, independence across mechanisms is preserved and each unobserved conditional response distribution is restricted to a capped probability simplex. Every collection of conditional distributions satisfying these local bounds is jointly attainable within a single compatible structural causal model. Optimizing the counterfactual expectation over the resulting product domain yields sharp attribution bounds for multistage interventions and mediated effects, and compatible response-vector polytopes extend these bounds to structural response restrictions. Exact algorithms for causal chains and general directed acyclic graphs, consistent plug-in endpoint estimators, and finite-sample and asymptotic confidence intervals are developed and illustrated using synthetic reactor experiments and health survey data.
\end{abstract}


\section{Introduction}

Retrospective causal attribution asks which earlier events contributed to an observed adverse outcome. When an industrial reactor shuts down following both a cooling fault and abnormal feed composition, evaluating whether repairing the cooling valve would have prevented that specific incident requires retrospective analysis beyond population-level intervention effects. Such questions belong to the causes-of-effects framework \citep{Pearl1999,TianPearl2000,DawidMusio2022}.

Observational distributions determine marginal potential-response probabilities at fixed input settings, whereas retrospective attribution depends on cross-world associations between potential responses under alternative inputs. In multistage systems, this unidentifiability propagates sequentially through intermediate variables. A complete factual record provides direct structural information by fixing the realized response of every constituent mechanism and restricting the conditional distributions of its remaining counterfactual responses.

Partial-identification bounds have been established for probabilities of causation and mediated attribution \citep{DawidHumphreysMusio2024,Zhang2025}, and posterior causal effects are point identified under binary monotonicity \citep{Lu2023}. General structural causal models also admit credal-network representations for counterfactual bounding \citep{Zaffalon2020}. Across these settings, marginal optimization yields sharp bounds whenever locally admissible conditional response distributions across distinct mechanisms and parent configurations can be realized simultaneously within a single compatible structural causal model.

In finite Markovian models under complete factual conditioning, mutual independence of exogenous disturbances preserves posterior independence across mechanisms and restricts each nonfactual response distribution to a capped probability simplex. Conditioning on the realized response at each node combines any collection of locally feasible distributions into a single compatible model. This joint attainability yields sharp attribution bounds for multistage interventions and mediated effects, and accommodates structural restrictions through compatible joint response distributions. The resulting product structure supports greedy backward recursion along causal chains, exact vertex enumeration across shared mechanisms in general directed acyclic graphs, and root-$n$ consistent plug-in estimation with finite-sample and asymptotic confidence intervals.

\section{Model and posterior distributions}
\label{sec:setup}

This section formulates the retrospective attribution target in a finite Markovian structural causal model and characterizes the conditional response distributions given a complete factual record.

\subsection{Model and attribution target}
\label{sec:three-node}

Let $V=(V_1,\ldots,V_d)$ be causally ordered variables with finite state spaces $\mathcal X_j$. Write $\mathcal V=\{V_1,\ldots,V_d\}$ and $\mathcal X_B=\prod_{V_j\in B}\mathcal X_j$ for $B\subseteq\mathcal V$. The parents of $V_j$ are $\Pa(V_j)\subseteq\{V_1,\ldots,V_{j-1}\}$, with configuration space $\mathcal P_j=\mathcal X_{\Pa(V_j)}$. Let $P_j=V_{\Pa(V_j)}$ be the random parent vector and $V_j(p)$ the potential response at $p\in\mathcal P_j$. The structural equations follow a Markovian causal model.

\begin{assumption}
\label{ass:markov}
For $j=1,\ldots,d$, $V_j=f_j(P_j,U_j)$, where the exogenous disturbances $U_1,\ldots,U_d$ are mutually independent.
\end{assumption}

Under Assumption~\ref{ass:markov}, the response vectors $R_j=\{V_j(p)\}_{p\in\mathcal P_j}$ are mutually independent across $j=1,\ldots,d$. For every $p\in\mathcal P_j$ with $\pr(P_j=p)>0$, consistency gives
\[
 \mu_j(w;p)=\pr\{V_j(p)=w\}=\pr(V_j=w\mid P_j=p).
\]
Write $\mu_j(p)=\{\mu_j(w;p):w\in\mathcal X_j\}$. Observational positivity, $\pr(P_j=p)>0$, is required only at parent configurations entering the attribution functional or the structural compatibility constraints introduced in \S\,\ref{sec:bounds}.

Fix a complete factual record $V=v$ with $\pr(V=v)>0$, let $p_j^\star=v_{\Pa(V_j)}$ denote the observed parent configuration of $V_j$, and take $Y=V_d$ as the outcome. For an intervention set $A\subseteq\mathcal V\setminus\{Y\}$ and level $a\in\mathcal X_A$, let $Y_a$ denote the counterfactual outcome under $\mathrm{do}(A=a)$. Consistency yields $Y_{v_A}=v_Y$ on $\{V=v\}$. For a prespecified score function $g:\mathcal X_Y\to\mathbb R$, define the posterior attribution effect
\begin{equation}
 \tau_g(A,a;v)=g(v_Y)-E\{g(Y_a)\mid V=v\}.
 \label{eq:multilevel-target}
\end{equation}
When $A=\{X\}$, write $\tau_g(a;v)=\tau_g(\{X\},a;v)$. For binary $X$ and $Y$ with $v_X=v_Y=1$, $a=0$, and $g(y)=y$, the target in \eqref{eq:multilevel-target} equals the posterior total causal effect $1-\pr(Y_0=1\mid V=v)$ \citep{Lu2023}.

Consider the binary chain $X\to M\to Y$ at the factual record $v=(1,1,1)$ under $\mathrm{do}(X=0)$. Write $u=\pr\{M(0)=1\mid V=v\}$ and $r=\pr\{Y(0)=1\mid V=v\}$. Conditional independence of the two response mechanisms and consistency $Y(1)=1$ yield
\[
 \pr(Y_0=1\mid V=v)=u+(1-u)r.
\]
Suppose $\pr(M=1\mid X=1)=0.6$, $\pr(M=1\mid X=0)=0.3$, $\pr(Y=1\mid M=1)=0.8$, and $\pr(Y=1\mid M=0)=0.2$. Dividing the joint probability bounds $\pr\{M(0)=1,M(1)=1\}\leq0.3$ and $\pr\{Y(0)=1,Y(1)=1\}\leq0.2$ by the factual probabilities $0.6$ and $0.8$ restricts $u$ to $[0,0.5]$ and $r$ to $[0,0.25]$. Joint attainability of these intervals yields $\pr(Y_0=1\mid V=v)\in[0,0.625]$ and $\tau_g(0;v)\in[0.375,1]$.

\subsection{Local posterior bounds and joint attainability}
\label{sec:local-constraints}
\label{sec:sharp}

Let $\alpha_j=\mu_j(v_j;p_j^\star)$ denote the factual response probability at node $j$, and write $q_j(w;p)=\pr\{V_j(p)=w\mid V=v\}$ with conditional probability vector $q_j(p)=\{q_j(w;p):w\in\mathcal X_j\}$. For any nonfactual parent configuration $p\neq p_j^\star$, the joint probability inequality $\alpha_jq_j(w;p)\leq\mu_j(w;p)$ restricts $q_j(p)$ to the capped probability simplex
\begin{equation}
 \mathcal K_j(p)=\{q\in\Delta(\mathcal X_j):\alpha_jq(w)\leq\mu_j(w;p),\ w\in\mathcal X_j\}.
 \label{eq:local-set}
\end{equation}
At the factual configuration, consistency fixes $\mathcal K_j(p_j^\star)=\{\delta_{v_j}\}$. Because the factual record $v$ remains fixed throughout, it is suppressed in the notation for feasible sets. The following theorem establishes posterior independence across mechanisms and the joint attainability of these local bounds.

\begin{theorem}
\label{thm:localization}
Under Assumption~\ref{ass:markov}, fix the marginal distributions $\{\mu_j(p)\}$ and a complete record $V=v$ with $\pr(V=v)>0$. Then:
\begin{itemize}
\item[(i)] $R_1,\ldots,R_d$ are mutually independent given $V=v$, with each conditional distribution determined by $V_j(p_j^\star)=v_j$;
\item[(ii)] the feasible set of $q_j(p)$ is $\mathcal K_j(p)$ for every $j$ and $p$;
\item[(iii)] every collection of conditional distributions $q_j(p)\in\mathcal K_j(p)$ across all $j$ and $p$ is jointly attained by a single compatible Markovian structural causal model.
\end{itemize}
\end{theorem}

The factual event $\{V=v\}=\bigcap_{j=1}^d\{V_j(p_j^\star)=v_j\}$ conditions each independent response vector on its own realized response, giving $\pr(V=v)=\prod_{j=1}^d\alpha_j$. When $\alpha_j<1$, any selection $q_j(p)\in\mathcal K_j(p)$ defines a residual distribution $r_j(p)=\{\mu_j(p)-\alpha_jq_j(p)\}/(1-\alpha_j)\in\Delta(\mathcal X_j)$ for $p\neq p_j^\star$. Taking the nonfactual responses to be conditionally independent given $V_j(p_j^\star)$, with conditional distribution $q_j(p)$ on $\{V_j(p_j^\star)=v_j\}$ and $r_j(p)$ on $\{V_j(p_j^\star)\neq v_j\}$, matches all marginal distributions and realizes the selected posterior distributions simultaneously (\S\,\ref{sec:sm-proof-thm1} of the Supplementary Material).

Conditioning on a partial factual record can induce dependence across mechanisms. In the binary chain $M\to Y$ with $M=U_M$, $Y(m)=I(m=U_Y)$, and independent $\operatorname{Bernoulli}(1/2)$ disturbances $U_M$ and $U_Y$, the event $\{Y=1\}=\{U_M=U_Y\}$ yields $\pr(U_M=1,U_Y=1\mid Y=1)=1/2\neq\pr(U_M=1\mid Y=1)\pr(U_Y=1\mid Y=1)=1/4$. Observing the complete record $(M,Y)=(1,1)$ fixes $U_M=U_Y=1$ and restores posterior independence.

\section{Sharp attribution bounds}
\label{sec:bounds}

Combining the local posterior sets across mechanisms yields sharp identification intervals for multistage interventions, mediated effects, and models subject to cross-world structural restrictions.

\subsection{Main result and binary cases}
\label{sec:binary}

For a collection of conditional distributions $q=\{q_j(p)\}$, define
\begin{equation}
 \Psi_g(q;A,a)=\sum_{\substack{z\in\mathcal X_{\mathcal V}:\\z_A=a}}
 g(z_Y)\prod_{j:V_j\notin A}q_j(z_j;z_{\Pa(V_j)}),
 \qquad
 \mathcal K_A=\prod_{j:V_j\notin A}\prod_{p\in\mathcal P_j}\mathcal K_j(p).
 \label{eq:psi-factorization}
\end{equation}
Theorem~\ref{thm:localization} gives $E\{g(Y_a)\mid V=v\}=\Psi_g(q;A,a)$. Write $L_g=L_g(A,a;v)$ and $U_g=U_g(A,a;v)$ for the endpoints
\begin{equation}
 [L_g,U_g]=\left[g(v_Y)-\max_{q\in\mathcal K_A}\Psi_g(q;A,a),\quad
                  g(v_Y)-\min_{q\in\mathcal K_A}\Psi_g(q;A,a)\right].
 \label{eq:sharp-endpoints}
\end{equation}
Joint attainability over the product domain $\mathcal K_A$ establishes the sharpness of these posterior attribution bounds.

\begin{theorem}
\label{thm:sharp}
Under Assumption~\ref{ass:markov}, if the marginal distributions entering \eqref{eq:psi-factorization} are identified and $\pr(V=v)>0$, the sharp identified interval for $\tau_g(A,a;v)$ is $[L_g(A,a;v),U_g(A,a;v)]$, and every value in this interval is attained by a compatible Markovian structural causal model.
\end{theorem}

The attainable expectations form a closed interval because $\mathcal K_A$ is compact and connected and $\Psi_g$ is continuous. In the binary system $X\to Y$ with $(X,Y)=(1,1)$, $a=0$, and $g(y)=y$, the effect $\tau_g(0;v)$ equals the probability of necessity, and \eqref{eq:local-set} yields the classical Fr\'echet bounds \citep{Pearl1999,TianPearl2000}:
\[
 \max\left\{0,1-\frac{\pr\{Y(0)=1\}}{\pr\{Y(1)=1\}}\right\}
 \leq\pr\{Y(0)=0\mid Y(1)=1\}
 \leq\min\left\{1,\frac{1-\pr\{Y(0)=1\}}{\pr\{Y(1)=1\}}\right\}.
\]
More generally, suppose $X$ and all its descendants are binary and each affected structural mechanism is coordinatewise nondecreasing in its affected parents. Every parent configuration $p\leq p_j^\star$ reached under $\mathrm{do}(X=0)$ then satisfies $V_j(p)\leq V_j(p_j^\star)$ almost surely. The conditional success probability $q_j(1;p)$ equals $0$ for $v_j=0$ and $\mu_j(1;p)/\alpha_j$ for $v_j=1$, recovering the point identification result of \citet{Lu2023}. When all observed variables equal one, this point-identified effect coincides with the unrestricted lower attribution bound.

\begin{corollary}
\label{cor:monotone-endpoint}
Suppose all variables are binary, the factual realization is $V=\boldsymbol1$, and the marginal distributions are compatible with coordinatewise nondecreasing structural mechanisms. For $A\subseteq\mathcal V\setminus\{Y\}$, $a=\boldsymbol0_A$, and $g(y)=y$, the point-identified attribution effect $\tau_{\mathrm{mono}}(A,a;\boldsymbol1)$ under structural monotonicity satisfies $\tau_{\mathrm{mono}}(A,a;\boldsymbol1)=L_g(A,a;\boldsymbol1)$.
\end{corollary}

\subsection{Mediated attribution}
\label{sec:mediated}

For a mediator subset $\mathcal S\subseteq\mathcal V\setminus(A\cup\{Y\})$, let $M=V_{\mathcal S}$ and write $M_a$ and $Y_{a,m}$ for the corresponding counterfactuals. Define the posterior total natural direct effect, pure natural indirect effect, and total natural indirect effect by
\[
\begin{aligned}
 \tau_g^{\mathrm{TNDE}}&=E\{g(Y_{v_A,M_{v_A}})-g(Y_{a,M_{v_A}})\mid V=v\},\\
 \tau_g^{\mathrm{PNIE}}&=E\{g(Y_{a,M_{v_A}})-g(Y_{a,M_a})\mid V=v\},\\
 \tau_g^{\mathrm{TNIE}}&=E\{g(Y_{v_A,M_{v_A}})-g(Y_{v_A,M_a})\mid V=v\}.
\end{aligned}
\]
Under conditional response distributions $q=\{q_j(p)\}$, consistency $M_{v_A}=v_M$ and $Y_{a,M_a}=Y_a$ yield
\[
\begin{aligned}
 \tau_g^{\mathrm{TNDE}}&=g(v_Y)-\Psi_g\{q;A\cup\mathcal S,(a,v_M)\},\\
 \tau_g^{\mathrm{PNIE}}&=\Psi_g\{q;A\cup\mathcal S,(a,v_M)\}-\Psi_g(q;A,a).
\end{aligned}
\]
In particular, $\tau_g^{\mathrm{TNDE}}=\tau_g\{A\cup\mathcal S,(a,v_M);v\}$. For a single mediator $M=V_m$, conditional independence gives
\[
 E\{g(Y_{v_A,M_a})\mid V=v\}
 =\sum_{w\in\mathcal X_m}\pr(M_a=w\mid V=v)
   \Psi_g\{q;A\cup\{V_m\},(v_A,w)\}.
\]
For multiple mediators, the same decomposition holds when no node in $\mathcal S$ is a descendant of any non-mediator descendant of $\mathcal S$ outside $A$ (\S\,\ref{sec:sm-mediated}, Lemma~\ref{lem:sm-multimediator-tnie}). For each mediated effect, minimizing and maximizing the corresponding functional over a common $q\in\mathcal K_A$ yields the sharp identified interval. Within each compatible model, $\tau_g(A,a;v)=\tau_g^{\mathrm{TNDE}}+\tau_g^{\mathrm{PNIE}}$, whereas the sharp interval endpoints of the two components are optimized separately over $\mathcal K_A$.

\subsection{Structural restrictions}
\label{sec:multilevel}

Cross-world shape restrictions constrain the response vector $R_j$ to a specified subset $\mathcal C_j\subseteq\prod_{p\in\mathcal P_j}\mathcal X_j$, coupling the conditional distributions $q_j(p)$ across distinct parent configurations $p\in\mathcal P_j$. Let $\mathcal Q_j$ denote the set of collections $\{q_j(p)\}_{p\in\mathcal P_j}$ induced by conditioning compatible joint response distributions supported on $\mathcal C_j$ on the factual event $\{V_j(p_j^\star)=v_j\}$ (\S\,\ref{sec:sm-multilevel-proofs}), and define the restricted product domain $\mathcal Q_A=\prod_{j:V_j\notin A}\mathcal Q_j$. In the unrestricted case $\mathcal C_j=\prod_{p\in\mathcal P_j}\mathcal X_j$, $\mathcal Q_j$ reduces to $\prod_{p\in\mathcal P_j}\mathcal K_j(p)$ and $\mathcal Q_A=\mathcal K_A$. Optimizing over $\mathcal Q_A$ yields sharp bounds under these structural restrictions.

\begin{theorem}
\label{thm:structural-sharp}
Suppose a compatible joint response distribution supported on $\mathcal C_j$ exists for every $j$. Under Assumption~\ref{ass:markov}, identified marginal distributions, and $\pr(V=v)>0$, \eqref{eq:sharp-endpoints} with $\mathcal Q_A$ in place of $\mathcal K_A$ gives the sharp identified interval for $\tau_g(A,a;v)$. Every value in this interval is attained by a compatible model satisfying the response restrictions.
\end{theorem}

For an ordered variable $V_j\in\{0,\ldots,M_j\}$ with an ordered parent $X\in\{x_0<\cdots<x_K\}$ and remaining parent set $S_j=\Pa(V_j)\setminus\{X\}$ with configurations $s\in\mathcal X_{S_j}$, bounded adjacent-increment restrictions take the form
\begin{equation}
 0\leq
 V_j(s,x_{k+1})-V_j(s,x_k)
 \leq b_{jk}
 \quad\text{almost surely},
 \qquad s\in\mathcal X_{S_j},\quad k=0,\ldots,K-1,
 \label{eq:adjacent-increment}
\end{equation}
for given integers $b_{jk}\geq 0$. Setting $b_{jk}=M_j$ imposes monotonicity across the ordered levels of $X$ at fixed $s$, whereas $b_{jk}=1$ restricts adjacent responses to unit increments. Write $F_j(r;s,x_k)=\pr(V_j\leq r\mid S_j=s,X=x_k)$, with $F_j(r;s,x_k)=0$ for $r<0$ and $F_j(r;s,x_k)=1$ for $r\geq M_j$. The following proposition characterizes marginal compatibility with these adjacent-increment bounds.

\begin{proposition}
\label{prop:cdf-sandwich}
Let $\mathcal C_j$ be the set of response vectors satisfying \eqref{eq:adjacent-increment}. A compatible joint response distribution exists if and only if
\begin{equation}
 F_j(r;s,x_k)
 \geq
 F_j(r;s,x_{k+1})
 \geq
 F_j(r-b_{jk};s,x_k)
 \qquad
 \text{for all }s,k,r.
 \label{eq:cdf-sandwich}
\end{equation}
\end{proposition}

Restrictions relative to the observed response take the form $V_j(p)\leq V_j(p_j^\star)$ for specified configurations $p\in\mathcal E_j\subseteq\mathcal P_j\setminus\{p_j^\star\}$ ordered below $p_j^\star$, with the inequality reversed for configurations ordered above $p_j^\star$. When these are the only comparisons, the conditional distributions can be optimized separately across parent configurations: bivariate distributions for each pair $\{V_j(p_j^\star),V_j(p)\}$ combine conditional on their common observed response, yielding $\mathcal Q_j=\prod_p\mathcal Q_j(p)$, where $\mathcal Q_j(p)$ is the set of conditional distributions admitted by that pairwise comparison (Lemma~\ref{lem:anchor-gluing}).

The five-node mediation network in \S\,\ref{sec:sm-multilevel}, comprising a binary covariate $C$, binary exposure $X$, three-level mediators $(M_1,M_2)$, and binary outcome $Y$ at $v=(1,1,3,3,1)$, illustrates these restrictions. Imposing monotonicity relative to the observed response on neither mechanism, the outcome alone, the mediators alone, and both mechanisms contracts the sharp interval for the posterior total causal effect $\tau_g(0;v)$ from $[0.200,1.000]$ to $[0.200,0.800]$, $[0.200,0.570]$, and $[0.200,0.452]$, respectively.

\section{Exact computation}
\label{sec:computation}

The multilinear structure of the counterfactual expectation over the product of capped simplices enables exact computation via backward recursion along causal chains and vertex enumeration in general directed acyclic graphs.

\subsection{Causal chains}
\label{sec:chain}

In the unrestricted model, $\Psi_g$ is linear in each conditional response distribution separately, so its extrema occur at products of vertices of $\mathcal K_j(p)$. Consider the causal chain $Z_0=X\to Z_1\to\cdots\to Z_m=Y$, where any additional parents of $Z_1,\ldots,Z_m$ are unaffected by $\mathrm{do}(X=a)$. Let $p_j(s)$ denote the parent configuration of $Z_j$ when $Z_{j-1}=s$ and all additional parents retain their factual values. Backward dynamic programming yields the sharp attribution bounds along the chain.

\begin{corollary}
\label{cor:chain}
Under Assumption~\ref{ass:markov}, identified marginal probabilities, and $\pr(V=v)>0$, suppose the affected variables form the chain above without additional response restrictions. Set $h_m^-=h_m^+=g$. For $j=m,\ldots,1$ and $s\in\mathcal X_{Z_{j-1}}$, define
\begin{equation}
 h_{j-1}^-(s)=\min_{q\in\mathcal K_{Z_j}\{p_j(s)\}}
              \sum_{z\in\mathcal X_{Z_j}}q(z)h_j^-(z),
 \qquad
 h_{j-1}^+(s)=\max_{q\in\mathcal K_{Z_j}\{p_j(s)\}}
              \sum_{z\in\mathcal X_{Z_j}}q(z)h_j^+(z).
 \label{eq:chain-recursion}
\end{equation}
The sharp interval for $\tau_g(a;v)$ is $[g(v_Y)-h_0^+(a),\,g(v_Y)-h_0^-(a)]$.
\end{corollary}

At a nonfactual configuration in \eqref{eq:chain-recursion}, sorting the states by their continuation values and allocating probability successively up to $\min\{1,\mu_{Z_j}[z;p_j(s)]/\alpha_{Z_j}\}$ yields the local optimum; ascending order attains the minimum and descending order the maximum. At the factual configuration, the feasible set is a singleton. Distinct predecessor states index distinct parent configurations, so Theorem~\ref{thm:localization}(iii) ensures that the optimizing choices are jointly attainable. With $L_j=|\mathcal X_{Z_j}|$, reusing the sorted state order at each stage requires $O\{\sum_{j=1}^m(L_j\log L_j+L_{j-1}L_j)\}$ operations. The explicit allocation rule and proof are given in \S\,\ref{sec:sm-proof-cor1}.

\subsection{General graphs}
\label{sec:dag-computation}

In a general directed acyclic graph, reconverging paths can cause distinct counterfactual branches to evaluate a mechanism $j$ at the same nonfactual parent configuration $p$, thereby sharing the conditional distribution $q_j(p)$. Let $\mathcal R_{\mathrm{act}}$ be the set of nonfactual pairs $(j,p)$ with $V_j\notin A$ and $p\neq p_j^\star$ entering $\Psi_g(q;A,a)$, and let $\mathcal K_j^{\mathrm{ext}}(p)$ denote the vertex set of $\mathcal K_j(p)$.

For a selected subset of mechanisms $\mathcal F\subseteq\mathcal V\setminus A$, partition the active nonfactual distributions as $q=(q_{\mathcal F},q_{-\mathcal F})$ according to whether $V_j\in\mathcal F$. Let $\mathcal K_{\mathcal F}^{\mathrm{ext}}$ and $\mathcal K_{-\mathcal F}^{\mathrm{ext}}$ denote the corresponding products of the vertex sets $\mathcal K_j^{\mathrm{ext}}(p)$. This decomposition reduces optimization over $\mathcal K_A$ to outer vertex enumeration over $\mathcal F$ combined with inner optimization over the remaining mechanisms.

\begin{proposition}
\label{prop:cutset}
Under the conditions of Theorem~\ref{thm:sharp}, for any selected $\mathcal F$ and $\operatorname{opt}\in\{\min,\max\}$,
\[
 \underset{q\in\mathcal K_A}{\operatorname{opt}}\Psi_g(q;A,a)
 =\underset{q_{\mathcal F}\in\mathcal K_{\mathcal F}^{\mathrm{ext}}}{\operatorname{opt}}\,
  \underset{q_{-\mathcal F}\in\mathcal K_{-\mathcal F}^{\mathrm{ext}}}{\operatorname{opt}}
  \Psi_g(q_{\mathcal F},q_{-\mathcal F};A,a),
\]
where factual conditional distributions remain fixed at $q_j(p_j^\star)=\delta_{v_j}$. Whenever a fixed outer selection leaves a non-negative weighted sum of components involving disjoint sets of free conditional distributions, those components can be optimized separately, with chain components evaluated via Corollary~\ref{cor:chain}.
\end{proposition}

\begin{figure}[ht]
\centering
\begin{tikzpicture}[>=Stealth, node distance=1.4cm, every node/.style={font=\normalsize}]
 \node[draw, circle, minimum size=0.82cm, inner sep=1pt] (A) at (0,0) {$A$};
 \node[draw, circle, minimum size=0.82cm, inner sep=1pt] (T) at (1.7,0) {$T$};
 \node[draw, circle, minimum size=0.82cm, inner sep=1pt] (P) at (3.4,0.65) {$P$};
 \node[draw, circle, minimum size=0.82cm, inner sep=1pt] (D) at (1.7,1.3) {$D$};
 \node[draw, circle, minimum size=0.82cm, inner sep=1pt] (Y) at (5.1,0) {$Y$};
 \draw[->, thick] (A) -- (T);
 \draw[->, thick, dashed] (T) -- (P);
 \draw[->, thick] (D) -- (P);
 \draw[->, thick] (T) -- (Y);
 \draw[->, thick, dashed] (P) -- (Y);
 \node[font=\small] at (2.55,-1.05) {(a) Reactor};

 \node[draw, circle, minimum size=0.82cm, inner sep=1pt] (X) at (7.5,0.3) {$X$};
 \node[draw, circle, minimum size=0.82cm, inner sep=1pt] (B) at (9.2,0.95) {$B$};
 \node[draw, circle, minimum size=0.82cm, inner sep=1pt, fill=gray!18] (C) at (9.2,-0.35) {$C$};
 \node[draw, circle, minimum size=0.82cm, inner sep=1pt] (Y2) at (10.9,0.3) {$Y$};
 \draw[->, thick] (X) -- (B);
 \draw[->, thick] (X) -- (C);
 \draw[->, thick] (B) -- (Y2);
 \draw[->, thick] (C) -- (Y2);
 \node[font=\small] at (9.2,-1.05) {(b) Diamond};
\end{tikzpicture}
\caption{(a) Reactor network: temperature $T$ indexes disjoint downstream conditional distributions along the direct (solid) and pressure-mediated (dashed) paths, so $\mathcal F=\emptyset$. (b) Diamond network: branches $B=0$ and $B=1$ share $q_C(0)$; shading marks the selected shared mechanism $\mathcal F=\{C\}$.}
\label{fig:dag-structures}
\end{figure}
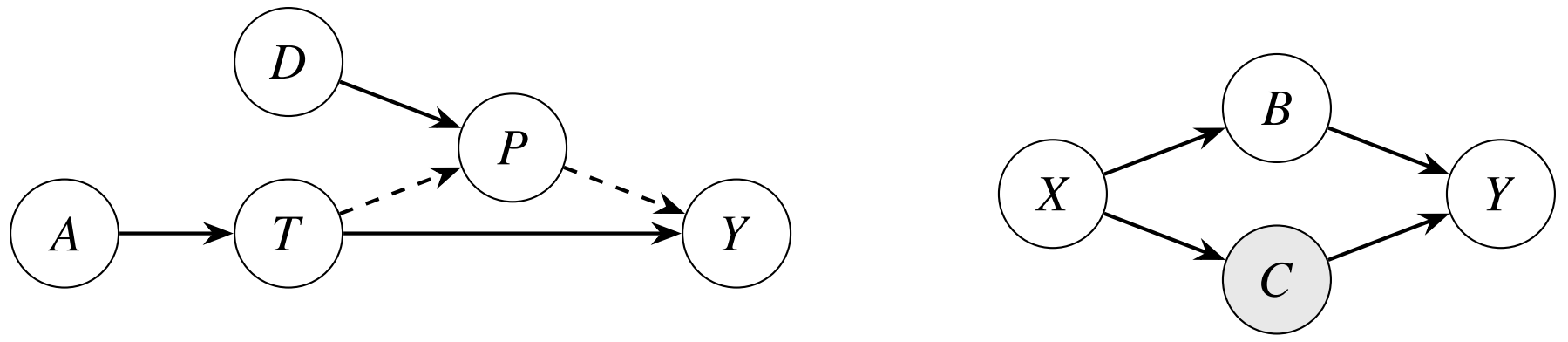

In the reactor graph of Fig.~\ref{fig:dag-structures}(a), recording $T$ in the parent sets of both $P$ and $Y$ separates all downstream conditional distributions across branches, permitting direct backward recursion with $\mathcal F=\emptyset$.

In the binary diamond network of Fig.~\ref{fig:dag-structures}(b) with $v=(1,0,0,1)$, $\pr\{B(1)=0\}=1$, $\pr\{B(0)=1\}=\pr\{C(1)=1\}=\pr\{C(0)=1\}=1/2$, and $Y(b,c)=I(b=c)$, both branches $b\in\{0,1\}$ of $B$ under $\mathrm{do}(X=0)$ share $q_C(0)\in\Delta(\{0,1\})$. Optimizing $q_C(0)$ separately across the two branches yields $[0,1]$ for $\pr(Y_0=1\mid V=v)$, whereas holding $q_C(0)$ fixed across branches gives $\{q_C(0;0)+q_C(1;0)\}/2=1/2$, which is recovered by selecting $\mathcal F=\{C\}$ (\S\,\ref{sec:sm-cutset-proof}).

\section{Estimation and inference}
\label{sec:estimation}

Replacing the population marginal distributions with sample estimates yields plug-in estimators for the sharp attribution bounds, whose polyhedral structure supports both finite-sample simultaneous confidence intervals and asymptotic directional inference.

\subsection{Estimation of sharp bounds}
\label{sec:plugin-stability}

Let $\mathcal D_A$ denote the feasible posterior domain, equal to $\mathcal K_A$ in the unrestricted model and $\mathcal Q_A$ under response restrictions. Replacing $\mu$ by an estimator $\widehat\mu$ defines the plug-in domain $\widehat{\mathcal D}_A$, over which optimizing the endpoint functionals yields $(\widehat L_g,\widehat U_g)$. In the unrestricted model, empirical or survey-weighted conditional frequencies give estimates in the product of probability simplices.

For ordered parents, fused least-squares estimation with tuning parameter $\gamma_n\geq0$ penalizes $L_1$ differences between adjacent conditional distributions \citep{GneitingRaftery2007,Tibshirani2005}. Let $\mathcal M_j$ be the set of marginal probability tables admitted by joint response distributions on $\mathcal C_j$. Compatibility is enforced directly during constrained fused estimation or by projecting preliminary estimates $\widetilde\mu_j$ onto $\mathcal M_j$ with fixed positive-definite quadratic weights $W_j$ (\S\,\ref{sec:sm-fused-rate}). Under fixed positive parent-cell probabilities, the fused estimator satisfies $\|\widehat\mu_j-\mu_j\|_\infty=O_p(n^{-1/2}+\gamma_n)$ and differs from the unpenalized constrained estimator by $O_p(\gamma_n)$, achieving first-order equivalence when $\sqrt n\gamma_n\to0$ \citep{KnightFu2000} (Lemma~\ref{lem:fused-rate}).

For fixed $(A,a)$, $v$, $g$, and $\{\mathcal C_j\}$, let $\mathcal I$ be the prespecified set of $m=|\mathcal I|$ active scalar probabilities entering the target functional or coupled to those probabilities through the response restrictions $\mathcal C_j$. Write $\|\widehat\mu-\mu\|_{\mathcal I}=\max_{(j,p,w)\in\mathcal I}|\widehat\mu_j(w;p)-\mu_j(w;p)|$. For an unrestricted outcome mechanism and $g(y)=I(y=v_Y)$, collapsing outcome categories retains only $\mu_Y(v_Y;p)$; the equality of the resulting extrema is proved in \S\,\ref{sec:sm-plugin-proof}. The following theorem establishes the stability and convergence rate of the plug-in endpoint estimators.

\begin{theorem}
\label{thm:plugin-consistency}
Fix finite state spaces and response sets $\{\mathcal C_j\}$, with $\mu_j\in\mathcal M_j$ for every $j$. Let $\widehat\mu_j\in\mathcal M_j$ be obtained by constrained fused least squares or weighted quadratic projection. If $\|\widehat\mu-\mu\|_{\mathcal I}=o_p(1)$ and the factual probabilities of the mechanisms entering the target satisfy $\min_j\alpha_j\geq\epsilon>0$, then
\[
 \max\{|\widehat L_g(A,a;v)-L_g(A,a;v)|,
        |\widehat U_g(A,a;v)-U_g(A,a;v)|\}
 =O_p(\|\widehat\mu-\mu\|_{\mathcal I}),
\]
uniformly over finite collections of interventions and scores.
\end{theorem}

\subsection{Confidence intervals}
\label{sec:inference}

Confidence intervals for partially identified parameters are constructed to cover the entire identified interval $[L_g,U_g]$ at nominal level $1-\delta$ \citep{ImbensManski2004}. Under independent sampling, simultaneous Clopper--Pearson intervals for the $m$ active probabilities are formed at level $1-\delta/m$ per coordinate \citep{ClopperPearson1934}. Replacing $\alpha_j$ by its lower confidence limit $\alpha_j^L$ and $\mu_j(w;p)$ by its upper confidence limit $\mu_j^U(w;p)$ relaxes the local constraint in \eqref{eq:local-set} to $\alpha_j^L q(w)\leq\mu_j^U(w;p)$ for $p\neq p_j^\star$. Optimizing $\Psi_g(q;A,a)$ over the product of these enlarged simplices yields an outer simultaneous confidence interval $[\widetilde L_g,\widetilde U_g]$ with guaranteed finite-sample coverage (\S\,\ref{sec:sm-inference-proofs}).

\begin{proposition}
\label{prop:confidence-envelope}
Under independent and identically distributed sampling, correctly specified causal and response restrictions, and $\pr(V=v)>0$,
\[
 \pr\{[L_g,U_g]\subseteq[\widetilde L_g,\widetilde U_g]\}\geq1-\delta
\]
for every sample size. Simultaneous Gaussian or survey-design confidence regions for the active probabilities give asymptotic coverage at least $1-\delta$ by the same construction.
\end{proposition}

For asymptotic inference, let $\mathcal M=\prod_j\mathcal M_j$ and let $\Pi_{\mathcal M}^W$ project each block using $W=\operatorname{diag}(W_1,\ldots,W_d)$. Writing $L_g(A,a;v;\eta)$ and $U_g(A,a;v;\eta)$ for the sharp endpoints evaluated at compatible margins $\eta\in\mathcal M$, define the composite endpoint map $T(\eta)=\{L_g(A,a;v;\Pi_{\mathcal M}^W\eta),\,U_g(A,a;v;\Pi_{\mathcal M}^W\eta)\}^\T$. At parameters with positive active factual probabilities, this map is piecewise smooth and Hadamard directionally differentiable. Where its derivative is nonlinear, the standard plug-in bootstrap is inconsistent \citep[Theorem~3.1]{FangSantos2019}.

Let $\widetilde\mu$ be an unprojected preliminary estimate, and let $\mathbb G_n^*$ consistently approximate its centred Gaussian limit conditionally on the data. With $\varepsilon_n\to0$ and $\sqrt n\varepsilon_n\to\infty$, the directional difference quotient is
\[
 Z_n^*=\frac{T(\widetilde\mu+\varepsilon_n\mathbb G_n^*)-T(\widetilde\mu)}{\varepsilon_n}.
\]
Writing $(\widehat L_g,\widehat U_g)^\T=T(\widetilde\mu)$ and $Z_n^*=(Z_{L,n}^*,Z_{U,n}^*)^\T$, let $\widehat c_{1-\delta}$ denote the conditional $(1-\delta)$ quantile of $\max\{Z_{L,n}^*,-Z_{U,n}^*\}$ given the data. This yields the confidence interval $\widehat{\mathcal I}_{g,n}=[\widehat L_g-n^{-1/2}\max\{0,\widehat c_{1-\delta}\},\,\allowbreak\widehat U_g+n^{-1/2}\max\{0,\widehat c_{1-\delta}\}]$ (\S\,\ref{sec:sm-directional}; \citealp{HongLi2018}), whose asymptotic validity follows from directional differentiability of $T$.

\begin{proposition}
\label{prop:directional-inference}
Suppose the state spaces and response sets are finite, $\mu\in\mathcal M$ has positive factual probabilities for all mechanisms entering the target, and $\sqrt n(\widetilde\mu-\mu)\to_d\mathbb G_\mu\sim N(0,\Sigma_\mu)$. Then $T$ is Hadamard directionally differentiable at $\mu$ and $\sqrt n\{T(\widetilde\mu)-T(\mu)\}\to_d Z=(Z_L,Z_U)^\T=T_\mu'(\mathbb G_\mu)$. If the distribution function of $\max\{Z_L,-Z_U\}$ is continuous and strictly increasing at its $(1-\delta)$ quantile, $\varepsilon_n\to0$, $\sqrt n\varepsilon_n\to\infty$, and the conditional multiplier law consistently estimates $N(0,\Sigma_\mu)$, then
\[
 \lim_{n\to\infty}\pr\{[L_g,U_g]\subseteq\widehat{\mathcal I}_{g,n}\}\geq1-\delta.
\]
\end{proposition}

\section{Numerical illustrations}
\label{sec:numerical}

Synthetic reactor experiments and data from a national health survey illustrate exact bound computation, finite-sample estimation, and confidence interval performance.

\subsection{Reactor example}
\label{sec:finite-sample}

In the five-variable Markovian reactor system depicted in Fig.~\ref{fig:dag-structures}(a), a binary cooling-valve fault $A\in\{0,1\}$ affects temperature $T\in\{0,1,2\}$; temperature and an ordered feed deviation $D\in\{0,1,2\}$ affect binary pressure $P\in\{0,1\}$; and $(T,P)$ govern shutdown severity $Y\in\{0,1,2\}$. The population conditional distributions are specified in \S\,\ref{sec:sm-reactor}.

For the observed incident $v=(A,D,T,P,Y)=(1,2,2,1,2)$, which occurs with population probability $0.02268$, the posterior attribution effects $\tau_g(\{A\},0;v)$ and $\tau_g(\{D\},0;v)$ are evaluated using the severe-shutdown indicator score $g(y)=I(y=2)$. Because temperature $T$ belongs to the parent sets of both $P$ and $Y$, Proposition~\ref{prop:cutset} yields the exact bounds by backward recursion with $\mathcal F=\emptyset$, summarized in Table~\ref{tab:reactor-bounds}.

\begin{table}[ht]
\tbl{Sharp posterior attribution effects and mediated components in the synthetic reactor system}
{\begin{tabular}{lc}
Intervention or component & Sharp interval\\[2pt]
Cooling repair, total effect, $\mathrm{do}(A=0)$ & $[0.6737,\ 0.9722]$\\
\quad Total natural direct effect, $\tau_g^{\mathrm{TNDE}}$ & $[0.5397,\ 0.8333]$\\
\quad Pure natural indirect effect, $\tau_g^{\mathrm{PNIE}}$ & $[0.0688,\ 0.2359]$\\
\quad Total natural indirect effect, $\tau_g^{\mathrm{TNIE}}$ & $[0.2399,\ 0.4630]$\\
Correct feed deviation, $\mathrm{do}(D=0)$ & $[0.1235,\ 0.2160]$\\
Moderate feed deviation, $\mathrm{do}(D=1)$ & $[0.0593,\ 0.1358]$\\
Repair both faults, $\mathrm{do}(A=0,D=0)$ & $[0.7337,\ 1.0000]$
\end{tabular}}
\label{tab:reactor-bounds}
\begin{tabnote}
Intervals are marginally sharp under unrestricted responses.
\end{tabnote}
\end{table}

Cooling repair has a larger attribution effect than feed correction in every compatible model: its lower bound $0.6737$ exceeds the feed-correction upper bound $0.2160$ by $0.4577$. Table~\ref{tab:reactor-bounds} also gives the pressure-mediated components and joint-repair bounds. Three compatible models with the same observational distribution attain the exact lower endpoint $382/567\approx0.6737$, the independent-response value $19/25=0.7600$, and the exact upper endpoint $35/36\approx0.9722$ (\S\,\ref{sec:sm-reactor-derivation}).

\begin{figure}[ht]
\centering
\begin{tikzpicture}
\begin{groupplot}[
 group style={group size=2 by 1,horizontal sep=1.30cm},
 width=.43\textwidth,height=4.6cm,
 xmin=150,xmax=8500,xtick={500,2000,8000},xticklabels={500,2000,8000},
 xlabel={Sample size},tick label style={font=\scriptsize},
 label style={font=\small},
 legend style={font=\scriptsize,draw=none,fill=white,inner sep=1pt},
 every axis plot/.append style={thick,mark size=2pt}]
\nextgroupplot[ymin=.90,ymax=1.01,ytick={.90,.95,1.00},
 ylabel={Coverage},legend pos=south east]
\addplot[black!55,dashed,mark=o,mark options={solid}] coordinates {(500,1) (2000,1) (8000,1)};
\addlegendentry{Full Clopper--Pearson}
\addplot[black!75,densely dashed,mark=square*,mark options={solid}] coordinates {(500,1) (2000,1) (8000,1)};
\addlegendentry{Active Clopper--Pearson}
\addplot[black,solid,mark=triangle*] coordinates {(500,.9360) (2000,.9520) (8000,.9565)};
\addlegendentry{Directional}
\addplot[black,densely dotted,no marks,forget plot] coordinates {(150,.95) (8500,.95)};
\nextgroupplot[ymin=.27,ymax=.93,ytick={.3,.5,.7,.9},ylabel={Mean width}]
\addplot[black!55,dashed,mark=o,mark options={solid}] coordinates {(500,.8850) (2000,.5932) (8000,.4415)};
\addplot[black!75,densely dashed,mark=square*,mark options={solid}] coordinates {(500,.8515) (2000,.5703) (8000,.4289)};
\addplot[black,solid,mark=triangle*] coordinates {(500,.4014) (2000,.3615) (8000,.3335)};
\addplot[black,densely dotted,no marks] coordinates {(150,.298501) (8500,.298501)};
\end{groupplot}
\end{tikzpicture}
\caption{Coverage (left) and mean width (right) in 2000 replicates for full Clopper--Pearson ($m_{\mathrm{cal}}=26$), active Clopper--Pearson ($m=12$), and directional intervals. Clopper--Pearson coverage curves coincide at $1.000$; dotted lines mark nominal coverage and identification width.}
\label{fig:main-coverage}
\end{figure}
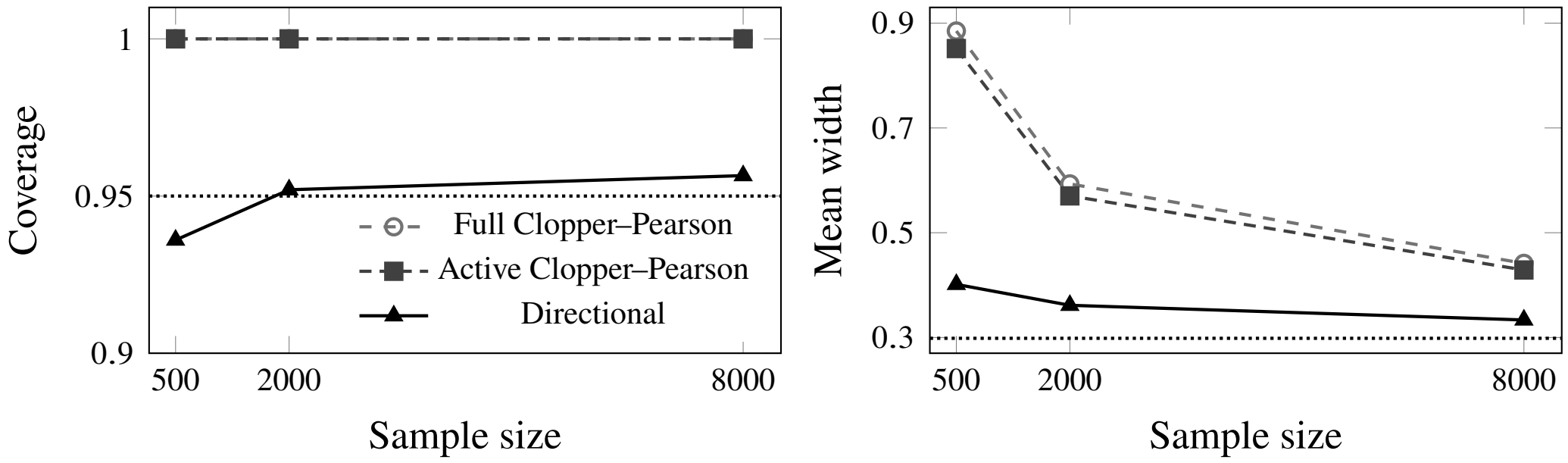

For cooling repair, $2000$ independent datasets are generated at each sample size $n\in\{500,2000,8000\}$ \citep{Morris2019}. Table~\ref{tab:mc-results} and Fig.~\ref{fig:main-coverage} compare simultaneous Clopper--Pearson confidence intervals calibrated over all $26$ model probabilities or the $12$ active probabilities with the numerical directional method. The multiplier and estimation settings are detailed in \S\,\ref{sec:sm-mc-protocol}.

\begin{table}[t]
\tbl{Coverage and excess width of nominal 95\% confidence intervals for the cooling-repair attribution interval across 2000 replicates}
{\begin{tabular}{rcccccc}
 & \multicolumn{2}{c}{Full Clopper--Pearson ($m_{\mathrm{cal}}=26$)} & \multicolumn{2}{c}{Active Clopper--Pearson ($m=12$)} & \multicolumn{2}{c}{Directional delta}\\
$n$ & Coverage & Excess width & Coverage & Excess width & Coverage & Excess width\\[2pt]
500  & 1.000 (0.000) & 0.587 (0.001) & 1.000 (0.000) & 0.553 (0.001) & 0.936 (0.006) & 0.103 (0.001)\\
2000 & 1.000 (0.000) & 0.295 (0.001) & 1.000 (0.000) & 0.272 (0.001) & 0.952 (0.005) & 0.063 (0.001)\\
8000 & 1.000 (0.000) & 0.143 (0.000) & 1.000 (0.000) & 0.130 (0.000) & 0.957 (0.005) & 0.035 (0.000)
\end{tabular}}
\label{tab:mc-results}
\begin{tabnote}
Coverage is the proportion of intervals containing the entire identified interval in Table~\ref{tab:reactor-bounds}. Excess width is mean interval width minus the population identification width $677/2268\approx0.2985$. Monte Carlo standard errors are in parentheses.
\end{tabnote}
\end{table}

Calibration over the active probabilities reduces interval length while retaining $1.000$ empirical coverage. The numerical directional method reduces excess width by $73\%$ to $81\%$ across the three sample sizes, with empirical coverage increasing from $0.936$ at $n=500$ to $0.952$ and $0.957$ at $n=2000$ and $8000$. At $n=500$, fused least-squares estimation reduces both the magnitude of upper-endpoint bias and the root mean squared errors of both endpoints, with estimation errors decreasing at larger sample sizes. Complete bias, root mean squared error, paired mean squared error, and tuning results are reported in \S\,\ref{sec:sm-mc-protocol}.

\subsection{Survey analysis}
\label{sec:real-data}

The 2017--2018 National Health and Nutrition Examination Survey provides complete records on $5530$ adults aged at least $20$ years across $15$ masked strata and $30$ primary sampling units \citep{Chen2020}. The analysis uses interview weights. The variables follow the working causal ordering $(X_1,X_2,X_3,Y)$, comprising recreational inactivity $X_1\in\{0,1\}$, ordered self-rated diet $X_2\in\{0,1,2\}$ from excellent or very good to fair or poor, coronary heart disease history $X_3\in\{0,1\}$, and hypertension diagnosis history $Y\in\{0,1,2\}$ with severe-outcome score $g(y)=I(y=2)$. The factual record $v=(X_1,X_2,X_3,Y)=(1,2,1,2)$ is observed in $42$ participants, representing a survey-weighted proportion of $0.0049$, and the smallest active outcome parent cell contains $23$ observations with a Kish effective sample size of $9.8$.

\begin{table}[t]
\setlength{\tabcolsep}{4pt}
\tbl{Survey-weighted posterior attribution intervals and 95\% confidence intervals in the 2017--2018 National Health and Nutrition Examination Survey}
{\begin{tabular}{lccc}
Candidate intervention & Active $m$ & Plug-in sharp interval & 95\% simultaneous interval\\[2pt]
Remove recreational inactivity, $\mathrm{do}(X_1=0)$ & 15 & $[0.013,\ 1.000]$ & $[0.000,\ 1.000]$\\
Improve self-rated diet, $\mathrm{do}(X_2=0)$ & 5 & $[0.123,\ 1.000]$ & $[0.000,\ 1.000]$\\
Remove coronary heart disease history, $\mathrm{do}(X_3=0)$ & 2 & $[0.508,\ 0.974]$ & $[0.166,\ 1.000]$\\
Joint lifestyle change, $\mathrm{do}(X_1=0,X_2=0)$ & 5 & $[0.407,\ 1.000]$ & $[0.000,\ 1.000]$
\end{tabular}}
\label{tab:nhanes-real}
\begin{tabnote}
Based on 5530 adults (42 matching $v=(1,2,1,2)$). Active $m$ is the number of prespecified scalar marginal coordinates entering each intervention's bound. For $\mathrm{do}(X_3=0)$, bivariate Fieller ratio inversion yields the 95\% confidence interval $[0.270,\ 1.000]$.
\end{tabnote}
\end{table}

Table~\ref{tab:nhanes-real} compares four structural interventions. The lifestyle interventions modify behavioural exposures, whereas $\mathrm{do}(X_3=0)$ sets the coronary heart disease indicator to zero prior to hypertension evaluation. Under $\mathrm{do}(X_3=0)$, $(X_1,X_2)=(1,2)$ remains fixed, and the sharp interval $[\max\{0,1-\beta_Y/\alpha_Y\},\,\allowbreak\min\{1,(1-\beta_Y)/\alpha_Y\}]$ depends only on $\alpha_Y=\pr(Y=2\mid X_1=1,X_2=2,X_3=1)$ and $\beta_Y=\pr(Y=2\mid X_1=1,X_2=2,X_3=0)$, estimated as $0.682$ and $0.336$ with plug-in interval $[0.508,0.974]$. Calibrating the simultaneous Gaussian region over $(\alpha_Y,\beta_Y)$ narrows the 95\% confidence interval from $[0.000,1.000]$ under calibration over all $54$ non-root probabilities to $[0.166,1.000]$, and Fieller inversion using the bivariate design covariance of $(\widehat\alpha_Y,\widehat\beta_Y)^\T$ for these two ratios yields $[0.270,1.000]$. The 95\% simultaneous confidence region intersects the compatible margin set defined by the dietary monotonicity and unit-increment inequalities in \eqref{eq:cdf-sandwich}; data processing, survey calibration, ratio inversion, and compatibility diagnostics are detailed in \S\,\ref{sec:sm-survey}.

\section{Discussion}

Conditioning on a complete factual record preserves independence across exogenous mechanisms in a Markovian structural causal model and reduces unrestricted nonfactual response constraints to explicit capped simplices. These feasible distributions combine through the realized response at each node into a single compatible model, establishing joint attainability and supporting exact backward recursion along causal chains and vertex enumeration across shared mechanisms in general directed acyclic graphs. Structural restrictions enter through compatible joint response distributions, whose polyhedral stability supports root-$n$ plug-in estimation and directional inference. Extending these results to partially observed intermediate records, dependent disturbances, or continuous responses requires optimization over coupled multi-node response laws or infinite-dimensional coupling sets.

\section*{Supplementary material}
\label{SM}
The Supplementary Material contains proofs, joint response constructions, algorithm details, estimation and inference formulas, complete simulation results, and additional mediation, survey, ordinal-exposure, and reliability examples.

\clearpage
\setcounter{section}{0}
\setcounter{equation}{0}
\setcounter{figure}{0}
\setcounter{table}{0}
\setcounter{lemma}{0}
\renewcommand{\thesection}{S\arabic{section}}
\renewcommand{\theequation}{S\arabic{equation}}
\renewcommand{\thefigure}{S\arabic{figure}}
\renewcommand{\thetable}{S\arabic{table}}
\renewcommand{\thelemma}{S\arabic{lemma}}
\renewcommand{\theHsection}{supp.\arabic{section}}
\renewcommand{\theHequation}{supp.\arabic{equation}}
\renewcommand{\theHfigure}{supp.\arabic{figure}}
\renewcommand{\theHtable}{supp.\arabic{table}}

\pdfbookmark[0]{Supplementary material}{supplement}
\begin{center}
  {\LARGE Supplementary material for ``\PaperTitle''\par}
  \vspace{1em}
  {\large \PaperAuthors\par}
\end{center}
\begin{abstract}
This supplementary material contains the technical proofs of all theoretical results in the main paper, details of the rules for selecting active probabilities, regularized margin estimators, and directional inference procedures, and complete derivations for the numerical and empirical studies. Additional examples on multilevel mediation, ordinal exposures, and reliability networks are also provided.
\end{abstract}

\section{Proofs of posterior identification}
\label{sec:sm-proofs}

\subsection{Proof of Theorem \ref{thm:localization}}
\label{sec:sm-proof-thm1}

\begin{proof}
The conditional independence in Theorem~\ref{thm:localization}(i) follows from nodewise exogenous independence.
Under Assumption~\ref{ass:markov}, the response vectors $R_1,\ldots,R_d$ are mutually independent.
The parent vector $P_j$ is a function of $R_1,\ldots,R_{j-1}$ and is therefore independent of $R_j$.
For each $p$ with $\pr(P_j=p)>0$, we have
\[
 \pr(V_j=w\mid P_j=p)
 =\pr\{V_j(p)=w\mid P_j=p\}
 =\pr\{V_j(p)=w\}
 =\mu_j(w;p),
\]
where the first equality holds by consistency and the second holds because $P_j$ is independent of $R_j$.
This proves the identification of $\mu_j(w;p)$.

For the observed evidence $V=v$, define $E_j=\{V_j(p_j^\star)=v_j\}$.
Then $\pr(E_j)=\alpha_j$.
By recursive substitution in the structural equations, we have
\[
 \{V=v\}=\bigcap_{j=1}^d E_j.
\]
Since each $E_j$ depends only on $R_j$, the mutual independence of $R_1,\ldots,R_d$ implies
\[
 \pr(V=v)=\prod_{j=1}^d\alpha_j.
\]
Because $\pr(V=v)>0$, each $\alpha_j$ is positive.
For any subsets $B_j$ of the response spaces, we then have
\[
 \pr(R_j\in B_j,\ j=1,\ldots,d\mid V=v)
 =\frac{\prod_{j=1}^d\pr(R_j\in B_j,E_j)}
        {\prod_{j=1}^d\pr(E_j)}
 =\prod_{j=1}^d\pr(R_j\in B_j\mid E_j),
\]
where the first equality follows from the independence of $R_1,\ldots,R_d$ and the fact that $E_j$ depends only on $R_j$.
Therefore, the response vectors are mutually independent conditional on $V=v$, as stated in part~(i).

By consistency, $q_j(p_j^\star)=\delta_{v_j}$.
For $p\ne p_j^\star$, part~(i) implies
\[
 \alpha_j q_j(w;p)
 =\pr\{V_j(p_j^\star)=v_j,V_j(p)=w\}
 \leq\mu_j(w;p).
\]
Thus $q_j(p)$ satisfies the constraints defining $\mathcal K_j(p)$ in \eqref{eq:local-set}.
To complete parts~(ii) and~(iii), we show that all admissible conditional distributions can be attained simultaneously by a single compatible Markovian structural causal model.

For each $j$ and $p$, choose $q_j(p)\in\mathcal K_j(p)$.
When $\alpha_j<1$, define
\[
 r_j(w;p)=\frac{\mu_j(w;p)-\alpha_j q_j(w;p)}{1-\alpha_j}
 \quad(p\ne p_j^\star).
\]
The constraints on $q_j(p)$ imply that $r_j(w;p)\geq0$ and $\sum_w r_j(w;p)=1$.
We construct a joint distribution for $R_j$ as follows.
First, let $V_j(p_j^\star)$ have marginal distribution $\mu_j(p_j^\star)$.
Conditional on $V_j(p_j^\star)=b$, let the remaining responses $\{V_j(p):p\ne p_j^\star\}$ be mutually independent, with
\[
 \pr\{V_j(p)=w\mid V_j(p_j^\star)=b\}
 =
 \begin{cases}
    q_j(w;p), & b=v_j,\\
    r_j(w;p), & b\ne v_j.
 \end{cases}
\]
By construction, for each $p\ne p_j^\star$,
\[
 \pr\{V_j(p)=w\}
 =\alpha_j q_j(w;p)+(1-\alpha_j)r_j(w;p)
 =\mu_j(w;p).
\]
Moreover, conditioning on $V_j(p_j^\star)=v_j$ yields $q_j(p)$ for every $p\ne p_j^\star$.
When $\alpha_j=1$, the constraints imply $q_j(p)=\mu_j(p)$.
In this case, we take the counterfactual responses to be mutually independent with these marginal distributions.

Finally, we take $R_1,\ldots,R_d$ to be mutually independent with their respective constructed distributions, and define $U_j=R_j$ and $f_j(p,U_j)=U_j(p)$.
The resulting structural causal model satisfies Assumption~\ref{ass:markov} and has marginal response distributions $\{\mu_j(p)\}$.
By part~(i), its conditional response distributions given $V=v$ equal the selected $q_j(p)$ for all $j$ and $p$ simultaneously.
This establishes attainability in part~(ii) and proves part~(iii).
\end{proof}

\subsection{Proof of Theorem \ref{thm:sharp}}
\label{sec:sm-proof-thm2}

\begin{proof}
Let $V_a$ denote the counterfactual vector under $\mathrm{do}(A=a)$; its conditional distribution given $V=v$ follows by causal recursion.
Under $\mathrm{do}(A=a)$, the variables in $A$ are fixed at $a$, and the remaining variables follow their structural equations.
By Theorem~\ref{thm:localization}(i), the response vector $R_j$ is independent of the preceding response vectors conditional on $V=v$.
Successive conditioning in the causal order therefore gives
\begin{equation}
 \pr(V_a=z\mid V=v)
 =1\{z_A=a\}\prod_{j:V_j\notin A}q_j(z_j;z_{\Pa(V_j)}).
 \label{eq:s-factorized-joint}
\end{equation}
If a preceding counterfactual event has probability zero, all its extensions also have probability zero, while the corresponding product vanishes by induction.
Multiplying \eqref{eq:s-factorized-joint} by $g(z_Y)$ and summing over $z$, we obtain $E\{g(Y_a)\mid V=v\}=\Psi_g(q;A,a)$.

To establish sharpness, note that $\mathcal K_A$ in \eqref{eq:psi-factorization} is the product of the sets $\mathcal K_j(p)$ over the nonintervened variables and their parent configurations.
By Theorem~\ref{thm:localization}(iii), every $q\in\mathcal K_A$ can be attained by a compatible model.
Moreover, $\mathcal K_A$ is compact and connected, and $\Psi_g(\cdot;A,a)$ is continuous.
Therefore, the set of attainable values of $E\{g(Y_a)\mid V=v\}$ is the closed interval between the minimum and maximum of $\Psi_g(\cdot;A,a)$ over $\mathcal K_A$.
By consistency, the attribution effect in \eqref{eq:multilevel-target} equals $g(v_Y)-E\{g(Y_a)\mid V=v\}$.
Subtracting the upper and lower endpoints, respectively, from $g(v_Y)$ completes the proof of Theorem~\ref{thm:sharp}.
\end{proof}

\subsection{Sharp bounds for posterior mediation effects}
\label{sec:sm-mediated}

For the posterior mediation effects in \S\,\ref{sec:mediated} of the main paper, let $\mathcal S$ denote the mediator nodes and $M=V_{\mathcal S}$ the corresponding vector. Consistency gives $M_{v_A}=v_M$ conditional on $\{V=v\}$ and $Y_{a,M_a}=Y_a$, so
\[
\begin{aligned}
 \tau_g^{\mathrm{TNDE}}&=g(v_Y)-\Psi_g(q;A\cup\mathcal S,(a,v_M)),\\
 \tau_g^{\mathrm{PNIE}}&=\Psi_g(q;A\cup\mathcal S,(a,v_M))-\Psi_g(q;A,a).
\end{aligned}
\]
Both counterfactual expectations are evaluated using the same $q\in\mathcal K_A$. The conditional response distributions of the mediators do not enter the first expectation, since their values are fixed at $v_M$.

For fixed values of the other conditional distributions, each expectation is affine in $q_j(p)$, because each term in the corresponding factorization contains at most one response probability from mechanism $j$. Their difference has the same property, so each objective attains its minimum and maximum at a product of vertices of the sets $\mathcal K_j(p)$. By Theorem~\ref{thm:localization}(iii), each such selection is attainable by a compatible model. Continuity of the objectives and connectedness of $\mathcal K_A$ then imply that every value between the respective endpoints is attainable. For the pure natural indirect effect, the difference is optimized over a common $q$; separate extrema of the two expectations can be attained by different distributions. This proves sharpness for both mediated effects.

For a single mediator $M=V_m$, with $\mathcal S=\{V_m\}$, consider the total natural indirect effect $\tau_g^{\mathrm{TNIE}}=g(v_Y)-E\{g(Y_{v_A,M_a})\mid V=v\}$.
The counterfactual $M_a$ depends only on the response vectors of $V_m$ and its ancestors.
On $\{V=v\}$, under $\mathrm{do}(A=v_A,V_m=w)$, every variable other than $V_m$ that is not a descendant of $V_m$ retains its factual value.
Hence $Y_{v_A,w}$ depends only on the response vectors of descendants of $V_m$.
By Theorem~\ref{thm:localization}(i), $M_a\perp Y_{v_A,w}\mid V=v$ for every $w\in\mathcal X_m$.
Consequently,
\begin{equation}
 E\{g(Y_{v_A,M_a})\mid V=v\}
 =\sum_{w\in\mathcal X_m}
   \pr(M_a=w\mid V=v)
   \Psi_g(q;A\cup\{V_m\},(v_A,w)).
 \label{eq:s-total-indirect}
\end{equation}
By \eqref{eq:s-factorized-joint}, each factor in every summand is affine in each conditional distribution separately.
Since the two factors involve disjoint sets of mechanisms, their product, and hence the sum, is also affine in each $q_j(p)$ separately.
\begin{lemma}[Block-ordered multimediator total natural indirect effect]
\label{lem:sm-multimediator-tnie}
Let $\mathcal S\subseteq\mathcal V\setminus(A\cup\{Y\})$, $M=V_{\mathcal S}$, and
\[
 \mathcal W=(\mathcal V\setminus(A\cup\mathcal S))\cap\operatorname{Desc}(\mathcal S).
\]
Here $\operatorname{Desc}(\mathcal S)$ denotes the set of descendants of nodes in $\mathcal S$. Suppose no node in $\mathcal S$ is a descendant of a node in $\mathcal W$.
Under Assumption~\ref{ass:markov} and complete factual conditioning, for every $m\in\mathcal X_{\mathcal S}$,
$M_a\perp Y_{v_A,m}\mid V=v$, and
\[
 E\{g(Y_{v_A,M_a})\mid V=v\}
 =\sum_{m\in\mathcal X_{\mathcal S}}\pr(M_a=m\mid V=v)
   \Psi_g(q;A\cup\mathcal S,(v_A,m)).
\]
This functional is affine in each conditional response distribution separately. Its extrema over $\mathcal K_A$ give the sharp interval for $\tau_g^{\mathrm{TNIE}}$ after subtraction from $g(v_Y)$.
\end{lemma}

\begin{proof}
The counterfactual $M_a$ is determined by response vectors at mediator nodes and their nonintervened ancestors.
The block-ordering condition places all these mechanisms outside $\mathcal W$.
Under $\mathrm{do}(A=v_A,M=m)$, nodes outside $\mathcal S\cup\mathcal W$ retain their factual values, so $Y_{v_A,m}$ depends only on response vectors indexed by $\mathcal W$.
Theorem~\ref{thm:localization}(i) gives the stated conditional independence and decomposition.
The two factors in each summand involve disjoint mechanisms and are separately affine in each conditional response distribution, so their product and sum have the same property.
Theorem~\ref{thm:localization}(iii) gives simultaneous attainment of the optimizing conditional distributions; continuity and connectedness of $\mathcal K_A$ give attainment of every intermediate value.
\end{proof}

\subsection{Binary monotonicity and proof of Corollary \ref{cor:monotone-endpoint}}
\label{sec:sm-monotone-endpoint}

The binary identification formula in \S\,\ref{sec:binary} of the main paper follows from the response restrictions.
Suppose $v_X=1$, the exposure and all its descendants are binary, and each affected structural mechanism is coordinatewise nondecreasing in its affected parents.
Under $\mathrm{do}(X=0)$, induction in the causal order shows that each counterfactual variable $V_{j,0}$ satisfies $V_{j,0}\leq v_j$.
For each affected node $V_j$ and reachable parent configuration $p\leq p_j^\star$, monotonicity implies $\{V_j(p)=1\}\subseteq\{V_j(p_j^\star)=1\}$.
Therefore, by Theorem~\ref{thm:localization}(i),
\[
 q_j(1;p)
 =\pr\{V_j(p)=1\mid V_j(p_j^\star)=v_j\}
 =
 \begin{cases}
  0,&v_j=0,\\
  \mu_j(1;p)/\alpha_j,&v_j=1.
 \end{cases}
\]

\begin{proof}
Let $\boldsymbol1$ and $\boldsymbol0$ denote vectors of ones and zeros of the required dimensions.
For the observed evidence $V=\boldsymbol1$ and intervention $\mathrm{do}(A=\boldsymbol0_A)$, let $q^\star$ collect the binary conditional distributions with $q_j^\star(1;p)=\mu_j(1;p)/\alpha_j$ for every $j$ and $p$.
Because the marginal distributions are compatible with nondecreasing mechanisms, $q_j^\star(1;p)$ is nondecreasing in $p$.
Moreover, the constraints on $\mathcal K_A$ imply $q_j(1;p)\leq q_j^\star(1;p)$ for every $q\in\mathcal K_A$.

To compare the counterfactual outcomes under $q$ and $q^\star$, use independent uniform random variables at the nonintervened nodes, with the same draw shared between the two constructions at each node.
Each response is one if this draw does not exceed the corresponding conditional success probability.
For parent configurations $p\leq p'$, we have
\[
 q_j(1;p)\leq q_j^\star(1;p)\leq q_j^\star(1;p').
\]
Thus, if the preceding variables under $q$ are no larger than those under $q^\star$, the same ordering holds at the current node.
Induction along the causal order gives the ordering of the outcomes and hence $\Psi_g(q;A,a)\leq\Psi_g(q^\star;A,a)$.

It remains to show that structural monotonicity identifies $q^\star$ and that $q^\star$ is attainable.
Since $V=\boldsymbol1$, every parent configuration satisfies $p\leq p_j^\star$.
Under structural monotonicity, $V_j(p)\leq V_j(p_j^\star)$ almost surely.
By Theorem~\ref{thm:localization}(i), every compatible monotone model therefore satisfies $q_j(1;p)=\mu_j(1;p)/\alpha_j=q_j^\star(1;p)$.
To establish attainability, let $U_1,\ldots,U_d$ be independent uniform random variables on $(0,1)$ and define $V_j(p)=1\{U_j\leq\mu_j(1;p)\}$.
These structural equations are nondecreasing in $p$, have the required marginal distributions, and induce $q^\star$ conditional on $V=\boldsymbol1$.
Thus structural monotonicity identifies the counterfactual success probability at its unrestricted maximum, so $\tau_{\mathrm{mono}}(A,a;\boldsymbol1)=L_g(A,a;\boldsymbol1)$.
\end{proof}

\section{Structural restrictions}
\label{sec:sm-multilevel-proofs}

\subsection{Proof of Theorem \ref{thm:structural-sharp}}

For the allowed response vectors $\mathcal C_j$ in the main paper, define
\begin{equation}
 \mathcal G_j
 =
 \left\{
 \pi_j\in\Delta(\mathcal C_j):
 \sum_{\substack{r\in\mathcal C_j:\\r(p)=w}}
 \pi_j(r)=\mu_j(w;p),
 \quad p\in\mathcal P_j,\;w\in\mathcal X_j
 \right\},
 \label{eq:response-law-polytope}
\end{equation}
assumed nonempty for each $j=1,\ldots,d$. Write $\mathcal G_j(\eta_j)$ for this set with marginal table $\eta_j$. Conditioning on $V_j(p_j^\star)=v_j$ updates $\pi_j\in\mathcal G_j$ to $\alpha_j^{-1}1\{r(p_j^\star)=v_j\}\pi_j(r)$ and induces the set of conditional response distributions
\[
 \mathcal Q_j
 =
 \left\{
 \{q_j(p)\}_{p\in\mathcal P_j}:
 q_j(w;p)
 =
 \frac{1}{\alpha_j}
 \sum_{\substack{
 r\in\mathcal C_j:\\
 r(p_j^\star)=v_j,\;r(p)=w
 }}
 \pi_j(r),
 \quad
 \pi_j\in\mathcal G_j
 \right\},
\]
with joint domain $\mathcal Q_A=\prod_{j:V_j\notin A}\mathcal Q_j$.

\begin{proof}
For any compatible model, let $\pi_j$ denote the distribution of $R_j$.
Then $\pi_j\in\mathcal G_j$ in \eqref{eq:response-law-polytope}.
By Theorem~\ref{thm:localization}(i), we have
\[
 \pr(R_j=r\mid V=v)
 =\alpha_j^{-1}1\{r(p_j^\star)=v_j\}\pi_j(r).
\]
Marginalizing this posterior distribution over the response coordinates yields a collection of conditional distributions in $\mathcal Q_j$.
By \eqref{eq:s-factorized-joint}, we therefore have $E\{g(Y_a)\mid V=v\}=\Psi_g(q;A,a)$ for some $q\in\mathcal Q_A$.

Conversely, consider any $q\in\mathcal Q_A$.
For each nonintervened variable, the definition of $\mathcal Q_j$ ensures that there exists $\pi_j\in\mathcal G_j$ inducing the selected $q_j$ through the posterior projection defining $\mathcal Q_j$.
For each intervened variable, choose any $\pi_j\in\mathcal G_j$.
We construct independent response vectors with these distributions and set $U_j=R_j$ and $f_j(p,U_j)=U_j(p)$.
The resulting structural causal model satisfies Assumption~\ref{ass:markov} and the response restrictions, and has the required marginal response distributions.
By Theorem~\ref{thm:localization}(i), it induces the selected $q$ conditional on $V=v$.

Each $\mathcal Q_j$ is a linear image of the nonempty compact convex polytope $\mathcal G_j$.
Thus $\mathcal Q_A$ is compact and connected.
As in the proof of Theorem~\ref{thm:sharp}, continuity of $\Psi_g(\cdot;A,a)$ implies that every value between its minimum and maximum is attained.
This proves Theorem~\ref{thm:structural-sharp}.
\end{proof}

\subsection{Proof of Proposition \ref{prop:cdf-sandwich}}

\begin{proof}
For a fixed mechanism $j$ and configuration $s$ of the remaining parents, write $W_k=V_j(s,x_k)$ for $k=0,\ldots,K$.
The distribution function of $W_k$ is $F_j(\cdot;s,x_k)$, as defined in the main paper.
Necessity follows directly from the response restrictions in \eqref{eq:adjacent-increment}.
The restrictions $0\leq W_{k+1}-W_k\leq b_{jk}$ imply
\[
 \{W_k\leq r-b_{jk}\}\subseteq
 \{W_{k+1}\leq r\}\subseteq
 \{W_k\leq r\}.
\]
Taking probabilities gives the inequalities in \eqref{eq:cdf-sandwich}.

For sufficiency, construct a joint distribution satisfying the response restrictions when these inequalities hold.
For $0<u<1$, define
\[
 Q_k(u)=\min\{r\in\{0,\ldots,M_j\}:F_j(r;s,x_k)\geq u\}.
\]
The inequalities in \eqref{eq:cdf-sandwich} imply $Q_k(u)\leq Q_{k+1}(u)\leq Q_k(u)+b_{jk}$.
Let $U$ be uniform on $(0,1)$ and set $W_k=Q_k(U)$ for every $k$.
Each $W_k$ then has distribution function $F_j(\cdot;s,x_k)$, and the adjacent-increment restrictions hold simultaneously.
Taking these quantile couplings independently across different $s$ gives a joint response distribution in $\mathcal G_j$, which proves sufficiency.
\end{proof}

\subsection{Restrictions relative to the observed response}
\label{sec:sm-anchor-gluing}

In \S\,\ref{sec:multilevel} of the main paper, each restricted response under restrictions relative to the observed response is compared only with the response at the observed parent configuration.
The following lemma shows that the resulting conditional distributions can be selected separately across parent configurations and attained simultaneously.

\begin{lemma}[Couplings with a common factual response]
\label{lem:anchor-gluing}
For mechanism $j$, suppose the only response restrictions are $V_j(p)\leq V_j(p_j^\star)$ for $p\in\mathcal E_j\subseteq\mathcal P_j\setminus\{p_j^\star\}$.
For each $p\ne p_j^\star$, choose a bivariate distribution for $\{V_j(p_j^\star),V_j(p)\}$ with margins $\mu_j(p_j^\star)$ and $\mu_j(p)$, supported on $\{w\leq z\}$ when $p\in\mathcal E_j$.
Then there exists a joint distribution of $R_j$ in $\mathcal G_j$ with all the chosen bivariate margins.
If $\alpha_j>0$, the corresponding conditional distributions given $V_j(p_j^\star)=v_j$ can therefore be selected separately and attained simultaneously, so that $\mathcal Q_j=\prod_{p\in\mathcal P_j}\mathcal Q_j(p)$.
\end{lemma}

\begin{proof}
Let $\gamma_p(z,w)$ denote the chosen bivariate probability mass function for $\{V_j(p_j^\star),V_j(p)\}$ at parent configuration $p\neq p_j^\star$.
Let $V_j(p_j^\star)$ have marginal distribution $\mu_j(p_j^\star)$.
Conditional on $V_j(p_j^\star)=z$ with $\mu_j(z;p_j^\star)>0$, let the remaining responses $V_j(p)$ ($p\ne p_j^\star$) be mutually independent, with
\[
 \pr\{V_j(p)=w\mid V_j(p_j^\star)=z\}
 =\frac{\gamma_p(z,w)}{\mu_j(z;p_j^\star)}.
\]
For values of $z$ with zero marginal probability, the conditional distributions can be chosen arbitrarily on $\{w\leq z\}$.
By construction, each pair $\{V_j(p_j^\star),V_j(p)\}$ has the chosen bivariate distribution $\gamma_p$.
Thus the joint distribution belongs to $\mathcal G_j$, satisfies all the restrictions relative to the observed response, and simultaneously attains the selected conditional distributions at $z=v_j$.
\end{proof}

\subsection{Joint identification of threshold probabilities}
\label{sec:sm-thresholds}

For an ordinal outcome with state space $\mathcal X_Y=\{0,\ldots,M_Y\}$, consider the threshold probabilities $\theta_r=\pr(Y_a<r\mid V=v)$, with the intervention and observed evidence fixed.
Let $\mathcal D_A$ denote the feasible posterior domain, equal to $\prod_{j:V_j\notin A}\prod_{p\in\mathcal P_j}\mathcal K_j(p)$ in the unrestricted model and $\prod_{j:V_j\notin A}\mathcal Q_j$ under response restrictions.
For each $q\in\mathcal D_A$, \eqref{eq:s-factorized-joint} gives
\[
 \theta_r
 =\sum_{\substack{z\in\mathcal X_{\mathcal V}:\\z_A=a,\ z_Y<r}}
     \prod_{j:V_j\notin A}q_j(z_j;z_{\Pa(V_j)})
 \quad(r=1,\ldots,M_Y).
\]
By Theorem~\ref{thm:localization}(iii) and Theorem~\ref{thm:structural-sharp}, varying $q$ over $\mathcal D_A$ gives exactly the jointly attainable values of $(\theta_1,\ldots,\theta_{M_Y})^\T$.
This set is compact and connected, as a continuous image of $\mathcal D_A$.
By summation by parts, any score $g$ satisfies
\[
 \tau_g(A,a;v)
 =g(v_Y)-g(M_Y)+\sum_{r=1}^{M_Y}\{g(r)-g(r-1)\}\theta_r,
\]
whose sharp bounds follow by minimizing and maximizing over a common $q\in\mathcal D_A$.

\section{Exact computation}
\label{sec:sm-computation}

\subsection{Proof of Corollary \ref{cor:chain}}
\label{sec:sm-proof-cor1}

\begin{proof}
Consider the chain $Z_0=X\to Z_1\to\cdots\to Z_m=Y$.
Let $S_j$ denote the parents of $Z_j$ other than $Z_{j-1}$.
By the assumptions of the corollary, these variables are unaffected by $\mathrm{do}(X=a)$ and thus retain their observed values.
For $Z_{j-1}=s$, the parent configuration of $Z_j$ is $p_j(s)=(s,v_{S_j})$, and the factual parent configuration is $p_{Z_j}^\star=p_j(v_{Z_{j-1}})$.

For the minimum, set $h_m^-=g$ and apply backward induction. If a common choice of subsequent conditional distributions attains $h_j^-(z)$ for every $z$, minimizing $\sum_zq(z)h_j^-(z)$ over $\mathcal K_{Z_j}(p_j(s))$ gives \eqref{eq:chain-recursion}, including the singleton factual case. Different states $s$ index distinct parent configurations. Theorem~\ref{thm:localization}(iii) makes all these minimizing choices jointly attainable, completing the induction. Replacing minima by maxima proves the recursion for $h^+$. Theorem~\ref{thm:sharp} gives attainment of every value between the endpoints.

For a fixed nonfactual configuration at stage $j$, write $h=h_j^-$, $\mu(z)=\mu_{Z_j}(z;p_j(s))$, and $\alpha=\alpha_{Z_j}$. To calculate the local minimum, order the $L_j$ states so that $h(z_{(1)})\leq\cdots\leq h(z_{(L_j)})$.
Assign probability successively in this order:
\[
 q^\star(z_{(k)})=\min\left\{\frac{\mu(z_{(k)})}{\alpha},
                  1-\sum_{\ell<k}q^\star(z_{(\ell)})\right\}
 \quad(k=1,\ldots,L_j).
\]
Since the upper bounds sum to $1/\alpha\geq1$, this allocation is feasible.
Moving probability mass from a state with a larger value of $h$ to a state with a smaller value whose upper cap has not been reached preserves feasibility and cannot increase the objective.
Repeating these transfers yields $q^\star$, which therefore attains the minimum; using $h=h_j^+$ and reversing the state order attains the maximum.
\end{proof}

\subsection{Proof of Proposition \ref{prop:cutset}, cutset selection, and shared-row analysis}
\label{sec:sm-cutset-proof}

For $\mathcal R_{\mathrm{act}}$ defined in \S\,\ref{sec:dag-computation}, the products of vertex sets are
\[
 \mathcal K_{\mathcal F}^{\mathrm{ext}}
 =\prod_{\substack{(j,p)\in\mathcal R_{\mathrm{act}}:\\V_j\in\mathcal F}}\mathcal K_j^{\mathrm{ext}}(p),
 \qquad
 \mathcal K_{-\mathcal F}^{\mathrm{ext}}
 =\prod_{\substack{(j,p)\in\mathcal R_{\mathrm{act}}:\\V_j\notin\mathcal F}}\mathcal K_j^{\mathrm{ext}}(p).
\]
At most one coordinate of a vertex is strictly between zero and its upper cap. The number of outer selections is $|\mathcal K_{\mathcal F}^{\mathrm{ext}}|=\prod_{(j,p)\in\mathcal R_{\mathrm{act}}:V_j\in\mathcal F}|\mathcal K_j^{\mathrm{ext}}(p)|$. Factual distributions are fixed and inactive distributions drop out of the objective.

\begin{proof}
For fixed values of the other conditional distributions, $\Psi_g(\cdot;A,a)$ is affine in $q_j(p)$, because each term in \eqref{eq:s-factorized-joint} contains at most one response probability from mechanism $j$.
Starting from a global optimizer, we can therefore replace $q_j(p)$ by a vertex in $\mathcal K_j^{\mathrm{ext}}(p)$ without changing the optimal value.
Repeating this replacement for each $(j,p)$ gives an optimizer at a product of vertices.
Partitioning the vertex selections according to whether $V_j$ belongs to $\mathcal F$ establishes the outer-inner vertex decomposition in Proposition~\ref{prop:cutset}.

For each fixed outer selection $q_{\mathcal F}\in\mathcal K_{\mathcal F}^{\mathrm{ext}}$, suppose the remaining objective is a sum of terms with fixed non-negative weights involving disjoint sets of free conditional distributions.
Then these terms can be optimized separately over their respective blocks of $q_{-\mathcal F}\in\mathcal K_{-\mathcal F}^{\mathrm{ext}}$.
When a component has the chain structure in Corollary~\ref{cor:chain}, its optimum is obtained by the greedy backward recursion.
If two components share a conditional distribution $q_j(p)$, that distribution must be optimized jointly across the components or its node included in $\mathcal F$.
\end{proof}

\medskip\noindent\textit{Deterministic selection of shared mechanisms.}\ 
To select $\mathcal F\subseteq\mathcal V\setminus A$ in a general directed acyclic graph under $\mathrm{do}(A=a)$:
\begin{enumerate}
\item[(1)] Identify the pairs $(j,p)\in\mathcal R_{\mathrm{act}}$ with $|\mathcal K_j^{\mathrm{ext}}(p)|>1$.
\item[(2)] Perform symbolic backward elimination in reverse topological order $j=d,d-1,\ldots,1$. If eliminating downstream nodes produces a continuation function at $V_j$ that depends on uneliminated variables outside $\Pa(V_j)$, compare the configurations of those variables. Whenever distinct configurations evaluate mechanism $j$ at the same parent configuration $p$ from step~(1), record $(j,p)$ as a shared pair and include $V_j$ in the candidate set of shared mechanisms.
\item[(3)] Among subsets $\mathcal F$ of candidate mechanisms such that, for every fixed $q_{\mathcal F}\in\mathcal K_{\mathcal F}^{\mathrm{ext}}$, the inner objective in Proposition~\ref{prop:cutset} decomposes into a non-negative weighted sum of components over pairwise disjoint sets of free conditional distributions, select one of minimum cardinality and then minimum outer vertex count $|\mathcal K_{\mathcal F}^{\mathrm{ext}}|$, breaking remaining ties by the topological order.
\end{enumerate}

In the diamond graph of Fig.~\ref{fig:dag-structures}(b), with edges $X\to B$, $X\to C$, $B\to Y$, and $C\to Y$, factual record $(X,B,C,Y)=(1,0,0,1)$, and deterministic outcome $Y(b,c)=1\{b=c\}$, eliminating $Y$ leaves the continuation value $h_C(b,x)=q_C(b;x)$ at node $C$, which depends on the uneliminated variable $B$. Both branches $b\in\{0,1\}$ share $q_C(0)\in\Delta(\{0,1\})$ at $x=0$. The marginal probabilities $\pr\{B(1)=0\}=1$, $\pr\{B(0)=1\}=1/2$, and $\pr\{C(1)=1\}=\pr\{C(0)=1\}=1/2$ give $q_B(0)=(1/2,1/2)^\T$ and
\[
 \pr(Y_0=1\mid V=v)=\tfrac12 q_C(0;0)+\tfrac12 q_C(1;0)=\tfrac12.
\]
Selecting $\mathcal F=\{C\}$ fixes $q_C(0)$ across both branches and recovers the exact value $1/2$.

\section{Estimation and inference}
\label{sec:sm-estimation-proofs}

\subsection{Probability estimation and regularization}
\label{sec:sm-fused-rate}

For a joint response law $\pi_j$, define $(B_j\pi_j)(w;p)=\sum_{r\in\mathcal C_j:r(p)=w}\pi_j(r)$. Then $\mathcal M_j=\{B_j\pi_j:\pi_j\in\Delta(\mathcal C_j)\}$, with $\mathcal M_j=\prod_p\Delta(\mathcal X_j)$ for unrestricted responses. Write $\mathcal M=\prod_j\mathcal M_j$, $W=\operatorname{diag}(W_1,\ldots,W_d)$, and $\Pi_{\mathcal M}^W=(\Pi_{\mathcal M_1}^{W_1},\ldots,\Pi_{\mathcal M_d}^{W_d})$.

Given a sample of size $n$, let $\ell\{\eta_j(p),w\}=\sum_{z\in\mathcal X_j}\{\eta_j(z;p)-1\{z=w\}\}^2$ denote the multicategory Brier loss for a candidate kernel $\eta_j=\{\eta_j(p):p\in\mathcal P_j\}$ \citep{GneitingRaftery2007}. When a parent coordinate $X$ has ordered levels $x_0<\cdots<x_K$ and remaining parents $s$, sparse parent cells are regularized through the adjacent-level fusion penalty $J_j(\eta_j)=\sum_s\omega_j(s)\sum_{k=0}^{K-1}\|\eta_j(s,x_{k+1})-\eta_j(s,x_k)\|_1$ with fixed weights $\omega_j(s)\geq 0$ \citep{Tibshirani2005}, setting $J_j(\eta_j)=0$ when no ordered parent is regularized. The constrained fused least-squares estimator is
\begin{equation}
 \widehat\mu_j\in\mathop{\arg\min}_{\eta_j\in\mathcal M_j}
 \left[\frac1n\sum_{i=1}^n\ell\{\eta_j(P_{ji}),V_{ji}\}
   +\gamma_n J_j(\eta_j)\right],
 \label{eq:fused-brier}
\end{equation}
where $P_{ji}$ is the observed parent configuration and $\widehat\mu_j^{(0)}$ is the unpenalized minimizer. In the unrestricted model it equals the empirical conditional frequencies. Compatibility can also be imposed by projection:
\begin{equation}
 \widehat\mu_j=\Pi_{\mathcal M_j}^{W_j}(\widetilde\mu_j)
 =\mathop{\arg\min}_{\eta_j\in\mathcal M_j}
 (\widetilde\mu_j-\eta_j)^\T W_j(\widetilde\mu_j-\eta_j),
 \qquad W_j\succ 0.
 \label{eq:margin-projection}
\end{equation}

\begin{lemma}[Rates and first-order equivalence of regularized margins]
\label{lem:fused-rate}
Suppose the state and parent spaces are fixed and finite, $\rho_j(p)=\pr(P_j=p)>0$ for all $p\in\mathcal P_j$, $\mathcal M_j$ is a correctly specified nonempty compact convex set, and the fusion weights $\omega_j(s)$ are fixed and finite. As $\gamma_n\to 0$, any measurable minimizer $\widehat\mu_j$ of \eqref{eq:fused-brier} and the unpenalized minimizer $\widehat\mu_j^{(0)}$ satisfy
\[
 \|\widehat\mu_j-\mu_j\|_\infty = O_p(n^{-1/2}+\gamma_n),
 \qquad
 \|\widehat\mu_j-\widehat\mu_j^{(0)}\|_\infty = O_p(\gamma_n).
\]
If $\sqrt{n}\gamma_n\to 0$, then $\sqrt{n}(\widehat\mu_j-\widehat\mu_j^{(0)})\to_p 0$, and in the unrestricted model $\sqrt{n}(\widehat\mu_j-\mu_j)\to_d N(0,\Sigma_j)$ with block-diagonal covariance $\Sigma_j^p = \rho_j(p)^{-1}\{\operatorname{diag}(\mu_j(p))-\mu_j(p)\mu_j(p)^\T\}$.
\end{lemma}

\begin{proof}
Fix a mechanism $j$ and write the empirical squared loss as
\[
 \ell_n(\eta_j)=\frac1n\sum_{i=1}^n\ell\{\eta_j(P_{ji}),V_{ji}\}.
\]
Expanding the squared loss gives $E\{\ell_n(\eta_j)-\ell_n(\mu_j)\}=\sum_{p\in\mathcal P_j}\rho_j(p)\|\eta_j(p)-\mu_j(p)\|_2^2$.
The empirical Hessian is diagonal, with entries $2N_j(p)/n$ repeated over the response categories, where $N_j(p)$ counts observations in parent cell $p$.
Since $N_j(p)/n\to_p\rho_j(p)>0$ for every $p$, the empirical loss is strongly convex with a constant bounded away from zero with probability tending to one.
Also, $\|\nabla\ell_n(\mu_j)\|_2=O_p(n^{-1/2})$, and the fixed finite weights ensure that $J_j$ is Lipschitz with constant $C<\infty$.

Comparing the penalized objective at $\widehat\mu_j$ and $\mu_j$ gives, with probability tending to one,
\[
 c\|\widehat\mu_j-\mu_j\|_2^2
 \leq
 \{\|\nabla\ell_n(\mu_j)\|_2+C\gamma_n\}
       \|\widehat\mu_j-\mu_j\|_2
\]
for a constant $c>0$.
Hence $\|\widehat\mu_j-\mu_j\|_\infty=O_p(n^{-1/2}+\gamma_n)$.

To compare the penalized and unpenalized estimators, note that both minimize their respective objectives over the same convex set $\mathcal M_j$.
Strong convexity and their optimality give, for a subgradient $h\in\partial J_j(\widehat\mu_j)$,
\[
 c\|\widehat\mu_j-\widehat\mu_j^{(0)}\|_2^2
 \leq
 (\widehat\mu_j-\widehat\mu_j^{(0)})^\T
 \{\nabla\ell_n(\widehat\mu_j)-\nabla\ell_n(\widehat\mu_j^{(0)})\}
 \leq \gamma_n h^\T(\widehat\mu_j^{(0)}-\widehat\mu_j)
 \leq C\gamma_n\|\widehat\mu_j-\widehat\mu_j^{(0)}\|_2.
\]
Therefore, $\|\widehat\mu_j-\widehat\mu_j^{(0)}\|_\infty=O_p(\gamma_n)$.
If $\sqrt{n}\gamma_n\to0$, the two estimators are first-order equivalent.

In the unrestricted model, all parent cells are nonempty with probability tending to one, and $\widehat\mu_j^{(0)}$ then equals the empirical conditional frequencies.
For each $p$ and $w$,
\[
 \sqrt{n}\{\widehat\mu_j^{(0)}(w;p)-\mu_j(w;p)\}
 =
 \frac1{\sqrt{n}}\sum_{i=1}^n
 \frac{1\{P_{ji}=p\}}{\rho_j(p)}
 \{1\{V_{ji}=w\}-\mu_j(w;p)\}
 +o_p(1).
\]
The multivariate central limit theorem and Slutsky's theorem yield the stated normal limit.
\end{proof}

\subsection{Proof of Theorem \ref{thm:plugin-consistency}}
\label{sec:sm-plugin-proof}

\begin{proof}
For fixed $v$, $g$, and $\{\mathcal C_j\}$, let $\mathcal P_{j,\mathrm{act}}(A,a)\subseteq\mathcal P_j$ contain the factual and reachable parent configurations of each mechanism entering the objective, together with all configurations coupled to them by $\mathcal C_j$; set $\mathcal P_{j,\mathrm{act}}(A,a)=\emptyset$ for the remaining mechanisms.
Define the active nonintervened mechanisms by
\[
 \mathcal J_{\mathrm{act}}=\{j:V_j\notin A\text{ and }\mathcal P_{j,\mathrm{act}}(A,a)\neq\emptyset\}.
\]

For an unrestricted outcome mechanism and $g(y)=1\{y=v_Y\}$, define the merged response $\widetilde Y(p)=1\{Y(p)=v_Y\}$ and $s_p=\mu_Y(v_Y;p)$.
For $p\neq p_Y^\star$, write $b_p=q_Y(v_Y;p)$ and $\alpha_Y=s_{p_Y^\star}$. Projection of $\mathcal K_Y(p)$ gives exactly
\[
 0\leq b_p\leq1,\qquad
 \alpha_Y b_p\leq s_p,\qquad
 \alpha_Y(1-b_p)\leq1-s_p.
\]
Because the upper bounds for $w\neq v_Y$ sum to $(1-s_p)/\alpha_Y$, they admit non-negative probabilities summing to $1-b_p$ if and only if $\alpha_Y(1-b_p)\leq1-s_p$, with explicit lift $q_Y(w;p)=(1-b_p)\mu_Y(w;p)/(1-s_p)$ for $w\neq v_Y$ when $s_p<1$, and $b_p=1$ when $s_p=1$.
At $p=p_Y^\star$, consistency fixes $b_p=1$. Theorem~\ref{thm:localization}(iii) realizes all lifted conditional distributions simultaneously.
Thus the original and merged models have identical extremal functionals, and only $s_p$ enters $\mathcal I$ for the outcome.

Since $\min_{j\in\mathcal J_{\mathrm{act}}}\alpha_j\geq\epsilon>0$ and $\|\widehat\mu-\mu\|_{\mathcal I}=o_p(1)$, the event $\min_{j\in\mathcal J_{\mathrm{act}}}\widehat\alpha_j\geq\epsilon/2$ has probability tending to one.
For an unrestricted mechanism $j\in\mathcal J_{\mathrm{act}}$, each active nonfactual set $\mathcal K_j(p)$ is determined directly by $\alpha_j$ and the active counterfactual probabilities in $\mathcal I$, using the merged binary response above for $Y$ with $g(y)=1\{y=v_Y\}$. Any omitted simplex-normalization probability is recovered by subtraction from one. On this event, the plug-in set $\widehat{\mathcal K}_j(p)$ has caps $\min\{1,\widehat\mu_j(w;p)/\widehat\alpha_j\}$ that are uniformly Lipschitz in these scalar coordinates. The simplex constraints are fixed, so the polyhedral error bound controls the Hausdorff distance:
\[
 d_H\{\widehat{\mathcal K}_j(p),\mathcal K_j(p)\}
 \leq C_j\|\widehat\mu-\mu\|_{\mathcal I}.
\]

For a mechanism $j\in\mathcal J_{\mathrm{act}}$ subject to cross-world restrictions $\mathcal C_j$, constraint closure places every restriction involving an active configuration wholly inside $\mathcal P_{j,\mathrm{act}}(A,a)$. Thus $\mathcal C_j=\mathcal C_{j,\mathrm{act}}\times\mathcal C_{j,-\mathrm{act}}$. Let $\eta_{j,\mathrm{act}}$ collect the complete marginal tables on this configuration set included in $\mathcal I$. With the fixed marginalization matrix $B_{j,\mathrm{act}}$,
\[
 \operatorname{proj}_{\mathrm{act}}\{\mathcal G_j(\eta_j)\}
 =\mathcal G_{j,\mathrm{act}}(\eta_{j,\mathrm{act}})
 =\{\pi\in\Delta(\mathcal C_{j,\mathrm{act}}):B_{j,\mathrm{act}}\pi=\eta_{j,\mathrm{act}}\}.
\]
Marginalization gives the forward inclusion. For the reverse inclusion, couple any law in $\mathcal G_{j,\mathrm{act}}(\eta_{j,\mathrm{act}})$ independently with the omitted-coordinate marginal of a law in the nonempty $\mathcal G_j(\eta_j)$. The product support ensures that this extension belongs to $\mathcal G_j(\eta_j)$.
Both \eqref{eq:fused-brier} and \eqref{eq:margin-projection} ensure that $\widehat\mu_{j,\mathrm{act}}$ and $\mu_{j,\mathrm{act}}$ belong to $B_{j,\mathrm{act}}\{\Delta(\mathcal C_{j,\mathrm{act}})\}$. Hoffman's error bound \citep{Hoffman1952} yields
\[
 d_H\{\mathcal G_{j,\mathrm{act}}(\widehat\mu_{j,\mathrm{act}}),\,
     \mathcal G_{j,\mathrm{act}}(\mu_{j,\mathrm{act}})\}
 \leq C_j\|\widehat\mu_{j,\mathrm{act}}-\mu_{j,\mathrm{act}}\|_\infty
 \leq C_j\|\widehat\mu-\mu\|_{\mathcal I}.
\]
On the same event, conditioning these joint laws on the observed response is uniformly Lipschitz because the factual probabilities are bounded below by $\epsilon/2$.
Consequently, the Hausdorff distance between the active projections of $\widehat{\mathcal D}_A$ and $\mathcal D_A$ satisfies
\[
 d_H\{\widehat{\mathcal D}_{A,\mathrm{act}},\,
     \mathcal D_{A,\mathrm{act}}\}
 =O_p\bigl(\|\widehat\mu-\mu\|_{\mathcal I}\bigr).
\]
Because $\Psi_g(\cdot;A,a)$ depends only on the active conditional response distributions in $\mathcal D_{A,\mathrm{act}}$ and is multilinear on a compact product of simplices, it is Lipschitz on that domain.
Its minimum and maximum therefore change by at most a constant multiple of $d_H\{\widehat{\mathcal D}_{A,\mathrm{act}},\mathcal D_{A,\mathrm{act}}\}$.
Taking the maximum over finitely many interventions and score functions proves the uniform convergence rate in Theorem~\ref{thm:plugin-consistency} and establishes that the endpoint map is locally Lipschitz on the subset of $\mathcal M$ with positive active factual probabilities.
\end{proof}

For finite samples, an empty parent cell $N_j(p)=0$ is assigned uniform empirical frequencies and the binomial confidence interval $[0,1]$. If an estimated active factual probability is zero or the confidence region $\mathcal H_{n,1-\delta}$ defined in \S\,\ref{sec:sm-inference-proofs} is empty, the reported interval is the logical range
\begin{equation}
 \left[g(v_Y)-\max_{w\in\mathcal X_Y}g(w),\
       g(v_Y)-\min_{w\in\mathcal X_Y}g(w)\right].
 \label{eq:s-logical-range}
\end{equation}

\subsection{Selection of active probabilities and proof of Proposition \ref{prop:confidence-envelope}}
\label{sec:sm-inference-proofs}

For an unrestricted outcome mechanism with indicator score, the simultaneous-interval construction uses the exact merged binary response in \S\,\ref{sec:sm-plugin-proof}. For each active coordinate $(j,p,w)\in\mathcal I$, let $[\ell_{j,p,w},u_{j,p,w}]$ be the two-sided Clopper--Pearson binomial confidence interval for $\mu_j(w;p)$ at level $1-\delta/m$ based on cell counts $N_j(w;p)$ and $N_j(p)$ \citep{ClopperPearson1934}. With $\mathcal M=\prod_{j=1}^d\mathcal M_j$, define the simultaneous confidence region $\mathcal H_{n,1-\delta}=\{\eta\in\mathcal M:\ell_{j,p,w}\leq \eta_j(w;p)\leq u_{j,p,w}\text{ for all }(j,p,w)\in\mathcal I\}$, and let $\alpha_j^L$ and $\mu_j^U(w;p)$ denote the minimum of $\eta_j(v_j;p_j^\star)$ and maximum of $\eta_j(w;p)$ over $\eta\in\mathcal H_{n,1-\delta}$. Define the enlarged local capped simplex $\widetilde{\mathcal K}_j(p_j^\star)=\{\delta_{v_j}\}$ and
\begin{equation}
 \widetilde{\mathcal K}_j(p)
 =
 \left\{
 q\in\Delta(\mathcal X_j):
 \alpha_j^L q(w)\leq\mu_j^U(w;p),\quad w\in\mathcal X_j
 \right\},
 \qquad p\neq p_j^\star.
 \label{eq:enlarged-capped-simplex}
\end{equation}
Minimizing and maximizing $\Psi_g(q;A,a)$ over the product $\widetilde{\mathcal D}_A=\prod_{j:V_j\notin A}\prod_{p\in\mathcal P_j}\widetilde{\mathcal K}_j(p)$ and subtracting from $g(v_Y)$ yields the outer simultaneous confidence interval $[\widetilde L_g,\widetilde U_g]$.

Table~\ref{tab:sm-active-counts} distinguishes the full-model calibration count $m_{\mathrm{cal}}$ from the intervention-specific count $m=|\mathcal I|$.
In the reactor model, the severe-shutdown score uses only $s_{tp}=\pr(Y=2\mid T=t,P=p)$ from the outcome mechanism. The full calibration set consists of the $2$ probabilities for $A$, $3$ for $D$, $6$ entries of $\mu_T(t;a)$, $9$ pressure success probabilities $\mu_P(1;t,d)$, and $6$ values $s_{tp}$: $m_{\mathrm{cal}}=2+3+6+9+6=26$.
For cooling repair, the active set consists of $\alpha_T=\mu_T(2;1)$, $\mu_T(1;0)$, $\mu_T(2;0)$, the three pressure probabilities at $D=2$, and all six $s_{tp}$: $m=3+3+6=12$.

In NHANES, each multinomial mechanism is parameterized by its full conditional probability vector, giving $m_{\mathrm{cal}}=6+12+36=54$ non-root coordinates and active counts $m\in\{15,5,2,5\}$ across the four interventions.

\begin{table}[t]
\tbl{Full calibration and intervention-specific active coordinate counts}
{\begin{tabular*}{\textwidth}{@{\extracolsep{\fill}}ll>{\raggedright\arraybackslash}p{6.9cm}c@{}}
Model & Scope & Marginal coordinates (counts) & Count\\[3pt]
Reactor & Full & $A$ (2), $D$ (3), $\mu_T$ (6), $\mu_P(1;\cdot)$ (9), $s_{tp}=\mu_Y(2;t,p)$ (6) & $m_{\mathrm{cal}}=26$\\[5pt]
 & $\mathrm{do}(A=0)$ & $\alpha_T,\mu_T(1;0),\mu_T(2;0)$ (3); $\mu_P(1;t,2)$ (3); $s_{tp}$ (6) & $m=12$\\[5pt]
NHANES & Full non-root & $X_2\mid X_1$ (6); $X_3\mid X_1,X_2$ (12); $Y\mid X_1,X_2,X_3$ (36) & $m_{\mathrm{cal}}=54$\\[5pt]
 & $\mathrm{do}(X_1=0)$ & $\alpha_{X_2},\mu_{X_2}(\cdot;0)$ (4); $\alpha_{X_3},\mu_{X_3}(1;0,x_2)$ (4); $\alpha_Y,\mu_Y(2;0,x_2,x_3)$ (7) & $m=15$\\[5pt]
 & $\mathrm{do}(X_2=0)$ & $\alpha_{X_3},\mu_{X_3}(1;1,0)$ (2); $\alpha_Y,\mu_Y(2;1,0,x_3)$ (3) & $m=5$\\[5pt]
 & $\mathrm{do}(X_3=0)$ & $\alpha_Y=\mu_Y(2;1,2,1)$; $\beta_Y=\mu_Y(2;1,2,0)$ & $m=2$\\[5pt]
 & $\mathrm{do}(X_1=0,X_2=0)$ & $\alpha_{X_3},\mu_{X_3}(1;0,0)$ (2); $\alpha_Y,\mu_Y(2;0,0,x_3)$ (3) & $m=5$
\end{tabular*}}
\label{tab:sm-active-counts}
\begin{tabnote}
Reactor indices are $t\in\{0,1,2\}$ and $p\in\{0,1\}$; NHANES indices are $x_2\in\{0,1,2\}$ and $x_3\in\{0,1\}$.
\end{tabnote}
\end{table}

\begin{proof}[of Proposition~\ref{prop:confidence-envelope}]
For independent and identically distributed observations, conditional on the parent-cell count $N_j(p)$, the count $N_j(w;p)$ follows a binomial distribution with success probability $\mu_j(w;p)$.
Using a Clopper--Pearson interval at level $1-\delta/m$ for each active probability in $\mathcal I$, the union bound gives $\pr(\mu\in\mathcal H_{n,1-\delta})\geq 1-\delta$ for every $n$ without requiring independence across coordinates.

On the event $\mu\in\mathcal H_{n,1-\delta}$, every feasible posterior distribution $q\in\mathcal D_A$ satisfies, for $p\neq p_j^\star$,
\[
 \alpha_j^L q_j(w;p)\leq\alpha_j q_j(w;p)
 \leq\mu_j(w;p)\leq\mu_j^U(w;p),
\]
and $q_j(p_j^\star)=\delta_{v_j}$, so $\mathcal D_A\subseteq\widetilde{\mathcal D}_A$ by \eqref{eq:enlarged-capped-simplex}.
Under response restrictions, $\mathcal D_A\subseteq\prod_{j:V_j\notin A}\prod_p\mathcal K_j(p)\subseteq\widetilde{\mathcal D}_A$, giving the same outer inclusion.
Minimizing and maximizing $\Psi_g(q;A,a)$ over $\widetilde{\mathcal D}_A$ therefore yields a simultaneous confidence interval $[\widetilde L_g,\widetilde U_g]$ containing $[L_g,U_g]$ with probability at least $1-\delta$.
Replacing the binomial confidence intervals by a simultaneous $(1-\delta)$ Gaussian or survey-design confidence region over $\mathcal I$ gives asymptotic coverage at least $1-\delta$ by the same inclusion argument.
\end{proof}

\subsection{Proof of Proposition \ref{prop:directional-inference} and implementation details}
\label{sec:sm-directional}

\begin{proof}
Write $\mathcal G_j(\eta_j)$ in standard form, including normalization, and remove redundant equalities. Each vertex is affine in $\eta_j$ on a polyhedral feasibility region, with finitely many bases. The counterfactual expectation is linear in each joint response law separately, so the endpoints are attained at products of such vertices. Substitution gives finitely many rational functions with denominators that are products of positive factual probabilities. They are continuously differentiable near $\mu$.

For a direction $h$ in the tangent cone of $\mathcal M$ at $\mu$, the feasible bases are constant along $\mu+th$ for sufficiently small $t>0$. Differences between candidate values have numerators that are either identically zero or have finitely many zeros near zero. Hence each endpoint follows a fixed differentiable candidate along this ray. The local Lipschitz bound in \S\,\ref{sec:sm-plugin-proof} makes these directional limits stable under converging directions: the difference of the corresponding quotients is bounded by $C\|h_n-h\|$.

The projection $\Pi_{\mathcal M}^W$ is piecewise affine and Hadamard directionally differentiable. Its derivative lies in the tangent cone. Composing this projection with the directional endpoint limits proves directional differentiability of $T$, with $T_\mu'(h)$ obtained by evaluating the endpoint derivative in the projected direction $(\Pi_{\mathcal M}^W)'_\mu(h)$. The directional delta method gives
\[
 \sqrt n\{T(\widetilde\mu)-T(\mu)\}\to_d
 Z=(Z_L,Z_U)^\T=T_\mu'(\mathbb G_\mu)
\]
\citep[Theorem~2.1]{FangSantos2019}.

For any compact set $K$ of directions, local Lipschitz continuity gives
\[
 \sup_{h\in K}\left\|
 \frac{T(\widetilde\mu+\varepsilon_nh)-T(\widetilde\mu)}{\varepsilon_n}
 -\frac{T(\mu+\varepsilon_nh)-T(\mu)}{\varepsilon_n}
 \right\|\leq\frac{2C\|\widetilde\mu-\mu\|}{\varepsilon_n}=o_p(1).
\]
Hadamard directional differentiability makes the second quotient converge uniformly on compact sets to $T_\mu'(h)$. Conditional tightness and consistency of $\mathbb G_n^*$ therefore give conditional weak convergence of $Z_n^*$ to $Z$ \citep{HongLi2018}. The stated quantile conditions yield consistency of $\widehat c_{1-\delta}$ and the coverage conclusion.
\end{proof}

\medskip\noindent\textit{Covariance estimation and multiplier implementation.}\ 
Write $(\widehat L_g,\widehat U_g)^\T=T(\widetilde\mu)$ and $Z_n^*=(Z_{L,n}^*,Z_{U,n}^*)^\T$. Let $\widehat c_{1-\delta}$ be the conditional $(1-\delta)$ quantile of $\max\{Z_{L,n}^*,-Z_{U,n}^*\}$ and set
\[
 \widehat{\mathcal I}_{g,n}=
 [\widehat L_g-n^{-1/2}\max\{0,\widehat c_{1-\delta}\},\,
  \widehat U_g+n^{-1/2}\max\{0,\widehat c_{1-\delta}\}].
\]

Under independent and identically distributed sampling, the asymptotic covariance matrix $\Sigma_\mu=\operatorname{diag}(\Sigma_1,\ldots,\Sigma_d)$ of the empirical conditional frequencies is block-diagonal across mechanisms $j$ and parent configurations $p\in\mathcal P_j$, because martingale-difference scores at distinct parent cells are uncorrelated.
We estimate each block by the plug-in multinomial covariance $\widehat\Sigma_j^p=\{N_j(p)/n\}^{-1}\{\operatorname{diag}(\widehat\mu_j(p))-\widehat\mu_j(p)\widehat\mu_j(p)^\T\}$ and generate independent multiplier draws $\mathbb G_n^*\sim N(0,\widehat\Sigma_\mu)$ conditionally on the sample.
When $\widehat\mu_j$ is obtained via fused least squares with tuning rate $\gamma_n=c n^{-3/4}$, Lemma~\ref{lem:fused-rate} ensures $\sqrt{n}(\widehat\mu_j-\widehat\mu_{j,\mathrm{emp}})\to_p 0$, so $\widehat\mu_j$ shares the same limiting distribution $N(0,\Sigma_\mu)$.
The endpoint map $T$ projects each perturbed vector $\widetilde\mu+\varepsilon_n\mathbb G_n^*$ onto $\mathcal M$ before evaluating the bounds. In the unrestricted model, each block is projected with Euclidean unit weights $W_j=I$ onto the product of simplices $\mathcal M_j=\prod_{p\in\mathcal P_j}\Delta(\mathcal X_j)$, and the final confidence limits in $\widehat{\mathcal I}_{g,n}$ are intersected with the logical range \eqref{eq:s-logical-range}.

\section{Numerical experiments}
\label{sec:sm-reactor}

\subsection{Data-generating distribution}

In the synthetic reactor system of \S\,\ref{sec:finite-sample}, the variables $A$ (cooling-valve fault) and $P$ (pressure) are binary, while $D$ (feed deviation), $T$ (temperature), and $Y$ (shutdown severity) take values in $\{0,1,2\}$, with directed edges $A\to T$, $T\to P$, $D\to P$, $T\to Y$, and $P\to Y$.

The root variables $A$ and $D$ are independent, with $\pr(A=1)=0.40$ and $(\pr(D=0),\pr(D=1),\pr(D=2))=(0.60,0.30,0.10)$.
The conditional distributions $\mu_T(t;a)=\pr(T=t\mid A=a)$ and $\mu_P(1;t,d)=\pr(P=1\mid T=t,D=d)$ are
\[
 \begin{array}{cccc}
 a&\mu_T(0;a)&\mu_T(1;a)&\mu_T(2;a)\\[2pt]
 0&0.70&0.20&0.10\\
 1&0.10&0.20&0.70
 \end{array}
 \qquad
 \begin{array}{cccc}
 t&d=0&d=1&d=2\\[2pt]
 0&0.10&0.17&0.25\\
 1&0.30&0.40&0.50\\
 2&0.65&0.78&0.90
 \end{array}.
\]
For the outcome, let $s_{tp}=\pr(Y=2\mid T=t,P=p)$ and $f_{tp}=0.25+0.10t+0.05p$, with
\[
 \begin{array}{ccc}
 t&s_{t0}&s_{t1}\\[2pt]
 0&0.05&0.25\\
 1&0.20&0.50\\
 2&0.50&0.90
 \end{array},
\]
and set $\pr(Y=0\mid T=t,P=p)=(1-s_{tp})(1-f_{tp})$ and $\pr(Y=1\mid T=t,P=p)=(1-s_{tp})f_{tp}$.
The factual record $V=v=(1,2,2,1,2)$ has probability $0.02268$.

\subsection{Sharp bounds and compatible response distributions}
\label{sec:sm-reactor-derivation}

For $g(y)=1\{y=2\}$ and observed outcome $v_Y=2$, $\tau_g(A,a;v)=\pr(Y_a<2\mid V=v)$ is the posterior probability of avoiding severe shutdown.
The factual probabilities are $\alpha_T=0.70$, $\alpha_P=0.90$, and $\alpha_Y=0.90$.
Let $b_{tp}=q_Y(2;t,p)$.
By Theorem~\ref{thm:localization}, its lower and upper bounds for $(t,p)\ne(2,1)$ are $b_{tp}^L=\max\{0,(s_{tp}-0.10)/0.90\}$ and $b_{tp}^U=\min\{1,s_{tp}/0.90\}$, with $b_{21}^L=b_{21}^U=1$ by consistency:
\[
 \begin{array}{ccccc}
 (t,p)&(0,0)&(0,1)&(1,0)&(1,1)\\[2pt]
 b_{tp}^L&0&1/6&1/9&4/9\\
 b_{tp}^U&1/18&5/18&2/9&5/9
 \end{array}
 \qquad
 \begin{array}{ccc}
 (t,p)&(2,0)&(2,1)\\[2pt]
 b_{tp}^L&4/9&1\\
 b_{tp}^U&5/9&1
 \end{array}.
\]
For pressure at $(t,d)\ne(2,2)$,
\[
\begin{aligned}
q_P^L(1;t,d)=\max\{0,(\mu_P(1;t,d)-0.10)/0.90\}
 &\leq q_P(1;t,d)\\
 &\leq\min\{1,\mu_P(1;t,d)/0.90\}=q_P^U(1;t,d),
\end{aligned}
\]
and $q_P(1;2,2)=1$.
At $D=2$, the minimum and maximum continuation values over $T=0,1,2$ are $(1/36,7/27,1)^\T$ and $(19/162,11/27,1)^\T$, respectively.

For cooling repair $\mathrm{do}(A=0)$, $q_T(1;0)\leq 2/7$ and $q_T(2;0)\leq 1/7$.
The three observationally equivalent compatible structural causal models referenced in \S\,\ref{sec:finite-sample} are constructed via Theorem~\ref{thm:localization}(iii) by setting the conditional response distributions as follows and completing the nonfactual residual distributions via $r_j(p)=\{\mu_j(p)-\alpha_j q_j(p)\}/(1-\alpha_j)$:
\begin{enumerate}
\item[(1)] \textit{Upper-endpoint model (SCM-U)}: Set $q_Y(2;t,p)=b_{tp}^L$, $q_P(1;t,2)=q_P^L(1;t,2)$ ($1/6$ at $t=0$, $4/9$ at $t=1$), and $q_T(0)=(1,0,0)^\T$. The severe-shutdown probability is $1/36$, attaining the upper attribution bound $U_g(\{A\},0;v)=35/36\approx 0.9722$.
\item[(2)] \textit{Lower-endpoint model (SCM-L)}: Set $q_Y(2;t,p)=b_{tp}^U$, $q_P(1;t,2)=q_P^U(1;t,2)$ ($5/18$ at $t=0$, $5/9$ at $t=1$), and $q_T(0)=(4/7,2/7,1/7)^\T$. The severe-shutdown probability is $(4/7)(19/162)+(2/7)(11/27)+1/7=185/567$, attaining the lower attribution bound $L_g(\{A\},0;v)=382/567\approx 0.6737$.
\item[(3)] \textit{Independent-response model (SCM-I)}: Take the potential responses $\{V_j(p):p\in\mathcal P_j\}$ within each mechanism $j$ to be mutually independent, so $q_j(p)=\mu_j(p)$ for all $p\neq p_j^\star$. This gives continuation values $(3/4)(1/20)+(1/4)(1/4)=1/10$ at $T=0$, $(1/2)(1/5)+(1/2)(1/2)=7/20$ at $T=1$, and $1$ at $T=2$; averaging over $q_T(0)=(7/10,2/10,1/10)^\T$ yields the exact attribution probability $1-\{(7/10)(1/10)+(2/10)(7/20)+1/10\}=19/25=0.7600$.
\end{enumerate}

For $\mathrm{do}(D=0)$ and $\mathrm{do}(D=1)$, temperature remains at $T=2$, with $q_P(1;2,0)\in[11/18,13/18]$ and $q_P(1;2,1)\in[34/45,13/15]$, yielding the sharp intervals $[10/81,35/162]\approx[0.1235,0.2160]$ and $[8/135,11/81]\approx[0.0593,0.1358]$.
For the joint repair $\mathrm{do}(A=0,D=0)$, the continuation values are $(0,5/27,127/162)^\T$ and $(13/162,1/3,71/81)^\T$, giving minimum shutdown probability $0$ and maximum $(4/7)(13/162)+(2/7)(1/3)+(1/7)(71/81)=151/567$, so the sharp interval is $[416/567,1]\approx[0.7337,1.0000]$.

\subsection{Posterior mediated effects through pressure}

Treating pressure $P$ as the mediator and setting $w_t=q_P(0;t,2)$, the bounds in \S\,\ref{sec:sm-reactor-derivation} give $w_0\in[13/18,5/6]$, $w_1\in[4/9,5/9]$, and $w_2=0$.
For the total natural direct effect, $E\{g(Y_{0,1})\mid V=v\}=\sum_{t=0}^2 q_T(t;0)b_{t1}\in[1/6,29/63]$, so $\tau_g^{\mathrm{TNDE}}\in[34/63,5/6]\approx[0.5397,0.8333]$.
For the pure natural indirect effect, $\tau_g^{\mathrm{PNIE}}=\sum_{t=0}^2 q_T(t;0)w_t(b_{t1}-b_{t0})$, with $b_{01}-b_{00}\in[1/9,5/18]$ and $b_{11}-b_{10}\in[2/9,4/9]$.
Hence $w_0(b_{01}-b_{00})\in[13/162,25/108]$ and $w_1(b_{11}-b_{10})\in[8/81,20/81]$, and optimizing over $q_T(0)$ yields
\[
 \tau_g^{\mathrm{PNIE}}\in\left[\frac67\frac{13}{162},\
         \frac57\frac{25}{108}+\frac27\frac{20}{81}\right]
 =[13/189,535/2268]\approx[0.0688,0.2359],
\]
attained at $q_T(0)=(6/7,0,1/7)^\T$ and $(5/7,2/7,0)^\T$, respectively.
For the total natural indirect effect, consistency under the factual exposure gives $T_1=2$, so \eqref{eq:s-total-indirect} reduces to $\tau_g^{\mathrm{TNIE}}=\{\sum_{t=0}^2 q_T(t;0)w_t\}(1-b_{20})\in[136/567,25/54]\approx[0.2399,0.4630]$.

\subsection{Simulation design and complete Monte Carlo results}
\label{sec:sm-mc-protocol}

For each sample size $n\in\{500,2000,8000\}$, we generate $2000$ independent datasets with random seed \texttt{20261014} and report Monte Carlo standard errors following \citet{Morris2019}.
The conditional distributions of $T$ given $A$ and $Y$ given $(T,P)$ are estimated by empirical frequencies, while $\pr(P\mid T,D)$ is estimated either by empirical frequencies or by five-fold cross-validated fused least squares with fixed penalty $J_P(\eta_P)=(1/3)\sum_{t=0}^2\sum_{d=0}^1\|\eta_P(t,d+1)-\eta_P(t,d)\|_1$ over tuning grids $\gamma_n=c n^{-\beta}$ with $\beta\in\{1/2,3/4\}$ and $c\in\{0,0.05,0.10,0.20,0.50\}$.
In each replicate, observations are partitioned randomly into five equal-sized folds; for each candidate $c$, the piecewise-quadratic fused least-squares objective \eqref{eq:fused-brier} is solved to a convergence tolerance of $10^{-10}$, the coefficient $\widehat c$ minimizing the held-out unpenalized Brier loss is selected, and $\pr(P\mid T,D)$ is refitted on the full sample at $\widehat\gamma_n=\widehat c n^{-\beta}$.
For directional inference, $800$ Gaussian multiplier draws $\mathbb G_n^*\sim N(0,\widehat\Sigma_\mu)$ with random seed \texttt{20280085} are generated using the plug-in multinomial covariance $\widehat\Sigma_\mu$ evaluated at the empirical frequencies.

The numerical directional intervals use $\varepsilon_n=n^{-1/3}$. At $n=500$, fusion with rate $n^{-3/4}$ reduces the lower- and upper-endpoint root mean squared errors from $4.19\times10^{-2}$ and $2.78\times10^{-2}$ to $4.05\times10^{-2}$ and $2.51\times10^{-2}$. At $n=2000$, its paired upper-endpoint mean squared error reduction is $7.29\times10^{-6}$ (Monte Carlo standard error $1.19\times10^{-6}$).

\begin{figure}[p]
\centering
\includegraphics[width=.75\linewidth]{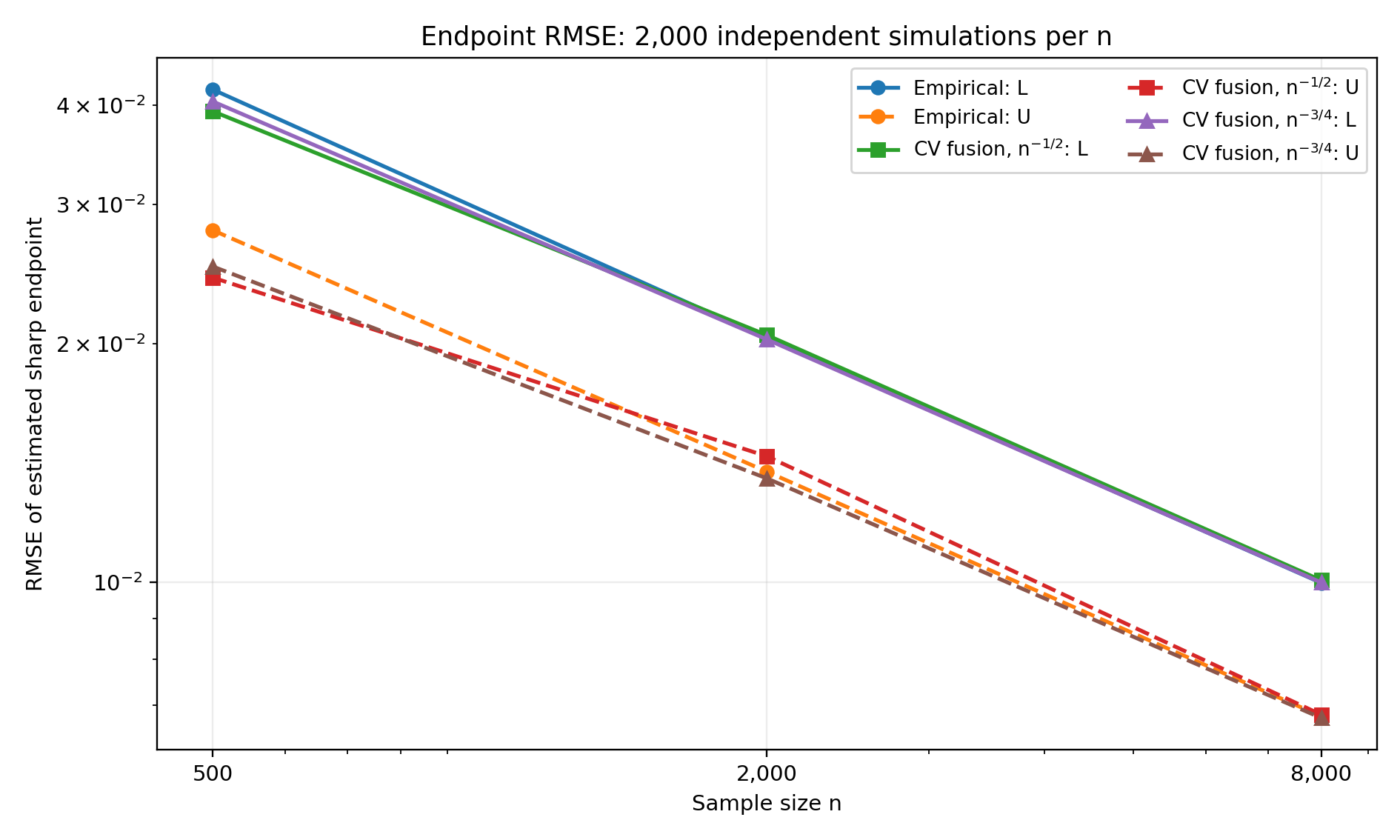}\par\smallskip
\includegraphics[width=.75\linewidth]{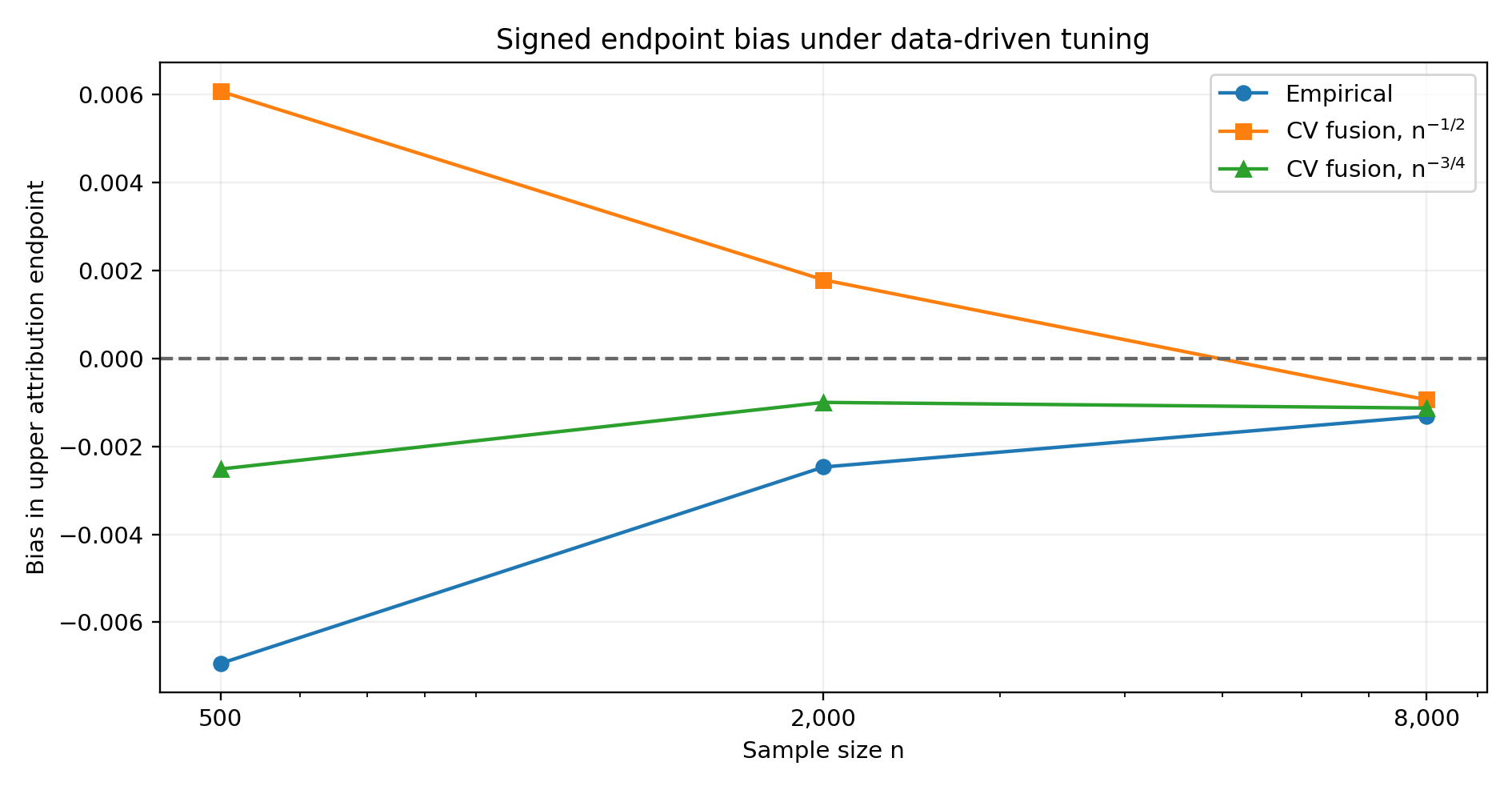}\par\smallskip
\includegraphics[width=.75\linewidth]{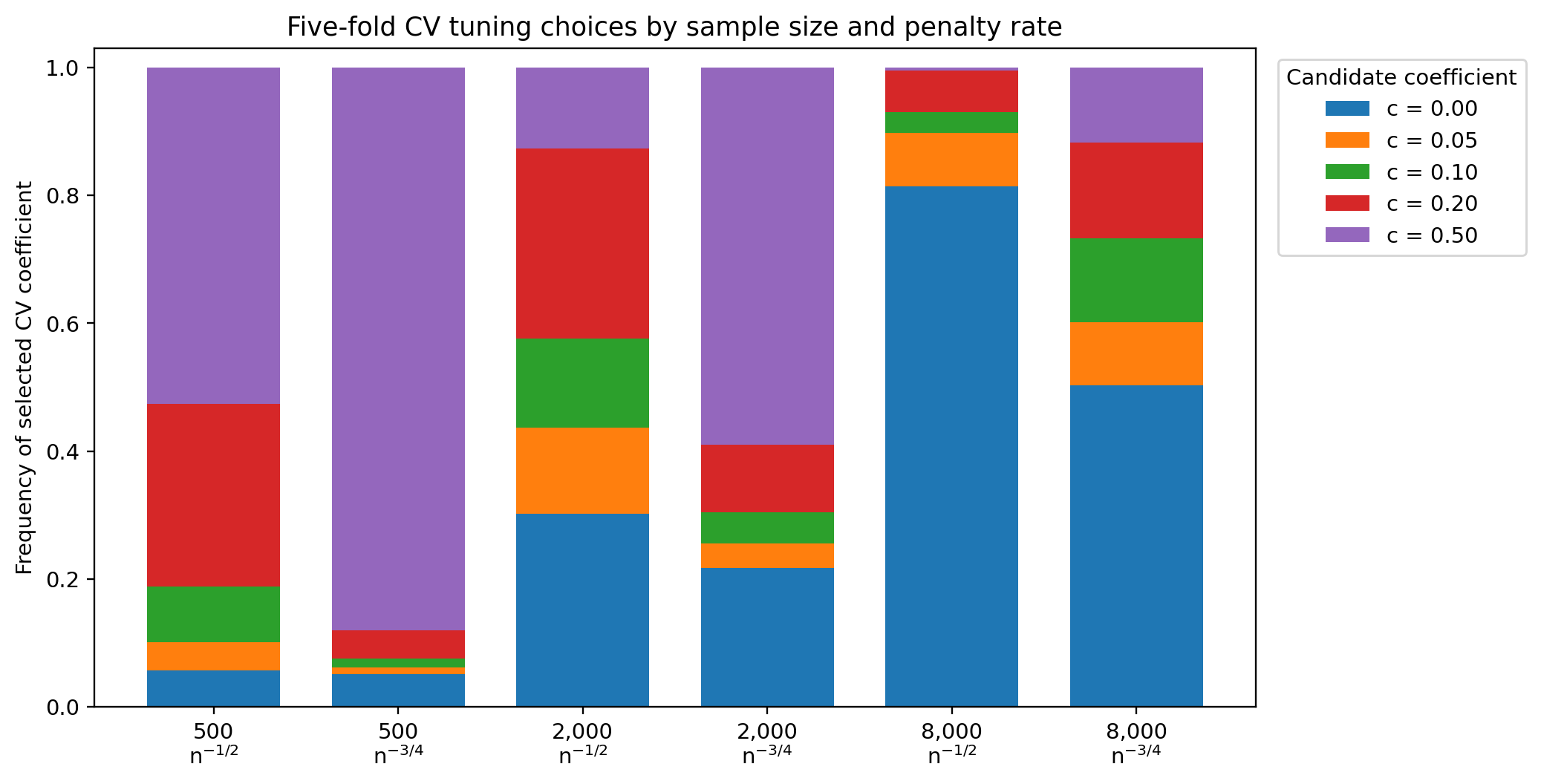}
\caption{Endpoint estimation and tuning in 2000 replicates. Top: root mean squared errors on a logarithmic scale. Middle: signed upper-endpoint bias. Bottom: five-fold cross-validation selection frequencies. Fusion uses penalty rates $n^{-1/2}$ and $n^{-3/4}$.}
\label{fig:sm-rmse-comparison}
\end{figure}

Tables~\ref{tab:sm-monte-carlo}--\ref{tab:sm-ci-diagnostics} and Fig.~\ref{fig:sm-rmse-comparison} present the endpoint estimation errors, paired mean squared error differences, cross-validation selection frequencies, and complete confidence interval diagnostics.

\begin{table}[!htbp]
\tbl{Endpoint estimation errors in 2000 replicates, multiplied by $100$}
{\begin{tabular*}{0.95\textwidth}{@{\extracolsep{\fill}}rlrrrrrr}
 & & \multicolumn{2}{c}{Bias} & \multicolumn{2}{c}{Standard deviation}
   & \multicolumn{2}{c}{Root mean squared error}\\
$n$ & Estimator & $L$ & $U$ & $L$ & $U$ & $L$ & $U$\\[2pt]
500 & Empirical & $-0.19$ & $-0.69$ & 4.19 & 2.70 & 4.19 & 2.78\\
500 & Fusion, $n^{-1/2}$ & 0.30 & 0.61 & 3.92 & 2.35 & 3.93 & 2.43\\
500 & Fusion, $n^{-3/4}$ & $-0.05$ & $-0.25$ & 4.05 & 2.49 & 4.05 & 2.51\\
2000 & Empirical & $-0.06$ & $-0.25$ & 2.03 & 1.36 & 2.03 & 1.38\\
2000 & Fusion, $n^{-1/2}$ & 0.12 & 0.18 & 2.05 & 1.43 & 2.05 & 1.44\\
2000 & Fusion, $n^{-3/4}$ & 0.01 & $-0.10$ & 2.03 & 1.35 & 2.03 & 1.35\\
8000 & Empirical & $-0.01$ & $-0.13$ & 1.00 & 0.67 & 1.00 & 0.68\\
8000 & Fusion, $n^{-1/2}$ & 0.01 & $-0.09$ & 1.01 & 0.67 & 1.01 & 0.68\\
8000 & Fusion, $n^{-3/4}$ & 0.00 & $-0.11$ & 1.00 & 0.67 & 1.00 & 0.67
\end{tabular*}}
\label{tab:sm-monte-carlo}
\begin{tabnote}
The columns $L$ and $U$ refer to the lower and upper endpoints.
Fusion coefficients are selected by five-fold cross-validation within the indicated grid.
Monte Carlo standard errors of the biases are at most $0.10$ on the displayed scale.
\end{tabnote}
\end{table}

\begin{table}[!htbp]
\tbl{Paired mean squared error differences for fusion with the $n^{-3/4}$ grid}
{\begin{tabular*}{0.95\textwidth}{@{\extracolsep{\fill}}rcc}
$n$ & Lower endpoint & Upper endpoint\\[2pt]
500 & $-11.73\ (1.01)$ & $-14.72\ (0.87)$\\
2000 & $-0.29\ (0.10)$ & $-0.73\ (0.12)$\\
8000 & $0.02\ (0.01)$ & $-0.05\ (0.01)$
\end{tabular*}}
\label{tab:sm-paired-errors}
\begin{tabnote}
Entries are the fusion mean squared error minus the empirical-frequency mean squared error, multiplied by $10^5$.
Monte Carlo standard errors of the paired differences are in parentheses.
Negative values favour fusion.
\end{tabnote}
\end{table}

\begin{table}[!htbp]
\tbl{Selection frequencies of the fusion coefficient $c$ under five-fold cross-validation in 2000 replicates}
{\begin{tabular*}{0.95\textwidth}{@{\extracolsep{\fill}}rlccccc}
$n$ & Penalty rate & $c=0$ & $c=0.05$ & $c=0.10$ & $c=0.20$ & $c=0.50$\\[2pt]
500  & $n^{-1/2}$ & 0.057 & 0.043 & 0.088 & 0.285 & 0.526\\
500  & $n^{-3/4}$ & 0.051 & 0.009 & 0.015 & 0.044 & 0.880\\
2000 & $n^{-1/2}$ & 0.302 & 0.135 & 0.140 & 0.297 & 0.126\\
2000 & $n^{-3/4}$ & 0.217 & 0.038 & 0.049 & 0.105 & 0.591\\
8000 & $n^{-1/2}$ & 0.815 & 0.083 & 0.033 & 0.065 & 0.005\\
8000 & $n^{-3/4}$ & 0.502 & 0.100 & 0.131 & 0.149 & 0.117
\end{tabular*}}
\label{tab:sm-cv-choices}
\begin{tabnote}
Entries are the proportions selecting each candidate coefficient $c$; their Monte Carlo standard errors are at most $0.012$.
\end{tabnote}
\end{table}

\begin{table}[!htbp]
\tbl{Coverage and width of nominal 95\% confidence intervals in 2000 replicates}
{\begin{tabular*}{0.95\textwidth}{@{\extracolsep{\fill}}rlccc}
$n$ & Procedure & Coverage & Mean width & Excess width\\[2pt]
500 & Full CP interval ($m_{\mathrm{cal}}=26$) & $1.0000\ (0.0000)$ & $0.8850\ (0.0012)$ & 0.5865\\
500 & Active CP interval ($m=12$) & $1.0000\ (0.0000)$ & $0.8515\ (0.0013)$ & 0.5530\\
500 & Directional delta & $0.9360\ (0.0055)$ & $0.4014\ (0.0011)$ & 0.1029\\
2000 & Full CP interval ($m_{\mathrm{cal}}=26$) & $1.0000\ (0.0000)$ & $0.5932\ (0.0007)$ & 0.2947\\
2000 & Active CP interval ($m=12$) & $1.0000\ (0.0000)$ & $0.5703\ (0.0007)$ & 0.2718\\
2000 & Directional delta & $0.9520\ (0.0048)$ & $0.3615\ (0.0005)$ & 0.0630\\
8000 & Full CP interval ($m_{\mathrm{cal}}=26$) & $1.0000\ (0.0000)$ & $0.4415\ (0.0003)$ & 0.1430\\
8000 & Active CP interval ($m=12$) & $1.0000\ (0.0000)$ & $0.4289\ (0.0003)$ & 0.1304\\
8000 & Directional delta & $0.9565\ (0.0046)$ & $0.3335\ (0.0002)$ & 0.0350
\end{tabular*}}
\label{tab:sm-ci-diagnostics}
\begin{tabnote}
Monte Carlo standard errors are in parentheses.
Coverage is the proportion of intervals containing $[382/567,35/36]$.
Excess width is mean width minus $0.298501$.
\end{tabnote}
\end{table}

\FloatBarrier
\section{Additional illustrations}
\label{sec:sm-additional}

\subsection{Multilevel mediation example}
\label{sec:sm-multilevel}

\begin{figure}[!htbp]
\centering
\includegraphics[width=.96\linewidth]{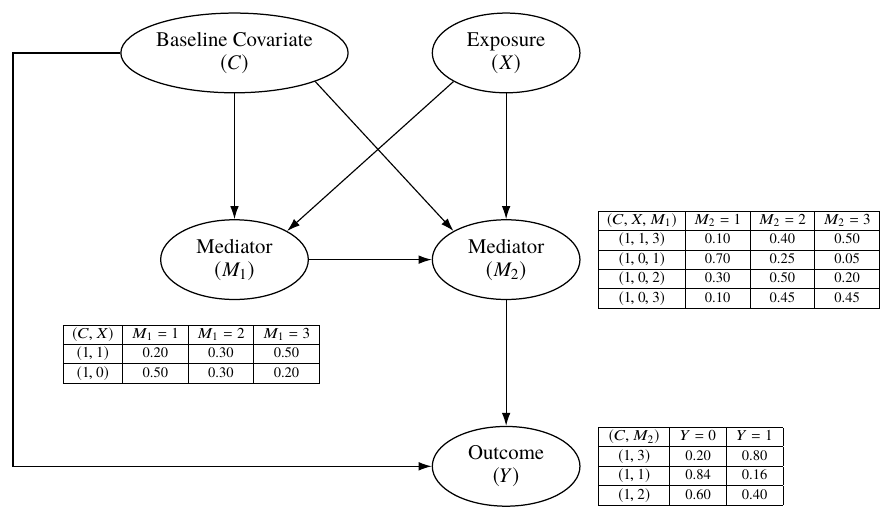}
\caption{Multilevel mediation network and conditional response probabilities for the factual record $(C,X,M_1,M_2,Y)=(1,1,3,3,1)$.}
\label{fig:multilevel-network}
\end{figure}

In the five-node multilevel mediation network summarized in \S\,\ref{sec:multilevel}, the directed edges in Fig.~\ref{fig:multilevel-network} are $C\to M_1$, $C\to M_2$, $C\to Y$, $X\to M_1$, $X\to M_2$, $M_1\to M_2$, and $M_2\to Y$, where $C,X,Y\in\{0,1\}$ and $M_1,M_2\in\{1,2,3\}$.
We condition on $v=(C,X,M_1,M_2,Y)=(1,1,3,3,1)$ and evaluate $\mathrm{do}(X=0)$, under which $C$ retains its observed value $1$.

Using parent orders $(C,X)$ for $M_1$, $(C,X,M_1)$ for $M_2$, and $(C,M_2)$ for $Y$, the required marginal response distributions are
\begin{align*}
 \mu_{M_1}(1,1)&=(0.20,0.30,0.50)^\T,&
 \mu_{M_1}(1,0)&=(0.50,0.30,0.20)^\T,\\
 \mu_{M_2}(1,1,3)&=(0.10,0.40,0.50)^\T,&
 \mu_{M_2}(1,0,1)&=(0.70,0.25,0.05)^\T,\\
 \mu_{M_2}(1,0,2)&=(0.30,0.50,0.20)^\T,&
 \mu_{M_2}(1,0,3)&=(0.10,0.45,0.45)^\T,\\
 \mu_Y(1,3)&=(0.20,0.80)^\T,&
 \mu_Y(1,1)&=(0.84,0.16)^\T,\\
 \mu_Y(1,2)&=(0.60,0.40)^\T,
\end{align*}
with factual probabilities $\alpha_{M_1}=0.50$, $\alpha_{M_2}=0.50$, and $\alpha_Y=0.80$.
By Theorem~\ref{thm:localization}, the unrestricted capped simplices with effective upper bounds $\min\{1,\mu_j(w;p)/\alpha_j\}$ are
\begin{align*}
 \mathcal K_{M_1}(1,0)
 &=\{q\in\Delta(\{1,2,3\}):q\leq(1.00,0.60,0.40)^\T\},\\
 \mathcal K_{M_2}(1,0,1)
 &=\{q\in\Delta(\{1,2,3\}):q\leq(1.00,0.50,0.10)^\T\},\\
 \mathcal K_{M_2}(1,0,2)
 &=\{q\in\Delta(\{1,2,3\}):q\leq(0.60,1.00,0.40)^\T\},\\
 \mathcal K_{M_2}(1,0,3)
 &=\{q\in\Delta(\{1,2,3\}):q\leq(0.20,0.90,0.90)^\T\},
\end{align*}
and $q_Y(1;1,1)\in[0,0.20]$, $q_Y(1;1,2)\in[0.25,0.50]$, and $q_Y(1;1,3)=1$.

In the unrestricted model, taking outcome probabilities at $(0,0.25,1)^\T$ gives minimum continuation values $(0,0.10,0.20)^\T$ for $m_1=1,2,3$ and minimum counterfactual outcome probability $0$ at $q_{M_1}(1,0)=(1,0,0)^\T$.
Taking outcome probabilities at $(0.20,0.50,1)^\T$ gives maximum continuation values $(0.43,0.70,0.95)^\T$ and maximum outcome probability $0.60(0.70)+0.40(0.95)=0.80$ at $q_{M_1}(1,0)=(0,0.60,0.40)^\T$, yielding $\tau_g(0;v)\in[0.200,1.000]$.
Under monotonicity relative to the observed response $W(p)\leq W(p_W^\star)$ imposed on the active nonfactual sets $\mathcal E_{M_1}=\{(1,0)\}$, $\mathcal E_{M_2}=\{(1,0,1),(1,0,2),(1,0,3)\}$, and $\mathcal E_Y=\{(1,1),(1,2)\}$, Lemma~\ref{lem:anchor-gluing} preserves separate optimization across parent configurations; the resulting sharp intervals and bivariate witness distributions are given in Tables~\ref{tab:multilevel} and~\ref{tab:s-witness}.

\begin{table}[!htbp]
\tbl{Sharp intervals in the multilevel mediation example under monotonicity relative to the observed response}
{\begin{tabular}{lccc}
Specification & Minimum continuation values & $\pr(Y_0=1\mid V=v)$ & $\tau_g(0;v)$\\[2pt]
Unrestricted & $(0.000,0.100,0.200)$ & $[0.000,0.800]$ & $[0.200,1.000]$\\
Outcome only & $(0.200,0.320,0.440)$ & $[0.200,0.800]$ & $[0.200,0.800]$\\
Mediators only & $(0.100,0.450,0.925)$ & $[0.430,0.800]$ & $[0.200,0.570]$\\
Mediators and outcome & $(0.280,0.580,0.950)$ & $[0.548,0.800]$ & $[0.200,0.452]$
\end{tabular}}
\label{tab:multilevel}
\begin{tabnote}
The maximum continuation values are $(0.430,0.700,0.950)$ under all four specifications, attained at $q_{M_1}(1,0)=(0,0.60,0.40)^\T$.
\end{tabnote}
\end{table}

\begin{table}[!htbp]
\tbl{Bivariate response distributions attaining the maximum and minimum unrestricted counterfactual risks}
{\begin{tabular}{lcc}
Mechanism and parent configuration & Maximum risk & Minimum risk\\[2pt]
$M_1:(1,0)$
& $\left(\begin{smallmatrix}0.20&0&0\\0.30&0&0\\0&0.30&0.20\end{smallmatrix}\right)$
& $\left(\begin{smallmatrix}0&0.20&0\\0&0.10&0.20\\0.50&0&0\end{smallmatrix}\right)$\\[3mm]
$M_2:(1,0,1)$
& $\left(\begin{smallmatrix}0.10&0&0\\0.40&0&0\\0.20&0.25&0.05\end{smallmatrix}\right)$
& $\left(\begin{smallmatrix}0.10&0&0\\0.10&0.25&0.05\\0.50&0&0\end{smallmatrix}\right)$\\[3mm]
$M_2:(1,0,2)$
& $\left(\begin{smallmatrix}0&0.10&0\\0.30&0.10&0\\0&0.30&0.20\end{smallmatrix}\right)$
& $\left(\begin{smallmatrix}0&0.10&0\\0&0.20&0.20\\0.30&0.20&0\end{smallmatrix}\right)$\\[3mm]
$M_2:(1,0,3)$
& $\left(\begin{smallmatrix}0&0.10&0\\0.10&0.30&0\\0&0.05&0.45\end{smallmatrix}\right)$
& $\left(\begin{smallmatrix}0&0.05&0.05\\0&0&0.40\\0.10&0.40&0\end{smallmatrix}\right)$\\[3mm]
$Y:(1,1)$
& $\left(\begin{smallmatrix}0.20&0\\0.64&0.16\end{smallmatrix}\right)$
& $\left(\begin{smallmatrix}0.04&0.16\\0.80&0\end{smallmatrix}\right)$\\[3mm]
$Y:(1,2)$
& $\left(\begin{smallmatrix}0.20&0\\0.40&0.40\end{smallmatrix}\right)$
& $\left(\begin{smallmatrix}0&0.20\\0.60&0.20\end{smallmatrix}\right)$
\end{tabular}}
\label{tab:s-witness}
\end{table}

\FloatBarrier
\subsection{NHANES analysis details}
\label{sec:sm-survey}

In the 2017--2018 NHANES analysis of \S\,\ref{sec:real-data}, we merge the files \texttt{DEMO\_J}, \texttt{PAQ\_J}, \texttt{DBQ\_J}, \texttt{MCQ\_J}, and \texttt{BPQ\_J} by participant identifier \texttt{SEQN} for adults aged at least $20$ years.
Recreational inactivity $X_1\in\{0,1\}$ sets $X_1=1$ for participants answering no (\texttt{2}) to both vigorous (\texttt{PAQ650}) and moderate (\texttt{PAQ665}) recreational activity, and $X_1=0$ otherwise.
Self-rated diet $X_2\in\{0,1,2\}$ from \texttt{DBQ700} codes excellent or very good (\texttt{1,2}) as $0$, good (\texttt{3}) as $1$, and fair or poor (\texttt{4,5}) as $2$.
Coronary heart disease history $X_3\in\{0,1\}$ from \texttt{MCQ160C} codes yes (\texttt{1}) as $1$ and no (\texttt{2}) as $0$.
Hypertension diagnosis history $Y\in\{0,1,2\}$ from \texttt{BPQ020} and \texttt{BPQ030} codes never told of hypertension (\texttt{BPQ020=2}) as $Y=0$, told only once (\texttt{BPQ020=1} and \texttt{BPQ030=2}) as $Y=1$, and told on two or more visits (\texttt{BPQ020=1} and \texttt{BPQ030=1}) as $Y=2$.
Participants with refused (\texttt{7}) or don't know (\texttt{9}) responses on any of these items are excluded, yielding the complete-case sample of $n=5530$ adults.

\medskip\noindent\textit{Survey variance and confidence calibration.}\ 
Survey-weighted marginal proportions $\widetilde\mu$ and their Taylor-linearized design covariance matrix $\widehat\Sigma_{\mathrm{surv}}$ are computed using interview weights \texttt{WTINT2YR}, $15$ masked variance strata (\texttt{SDMVSTRA}), and $30$ primary sampling units (\texttt{SDMVPSU}), providing $30-15=15$ design degrees of freedom \citep{Chen2020}.
For each intervention $(A,a)$ in Table~\ref{tab:sm-active-counts}, let $\widehat\Sigma_{\mathrm{act}}$ denote the $m\times m$ submatrix of $\widehat\Sigma_{\mathrm{surv}}$ on $\mathcal I$.
We generate $150{,}000$ draws from $N(0,\widehat\Sigma_{\mathrm{act}})$ with random seed \texttt{20261014} to calibrate the $95\%$ simultaneous critical value $c_{0.95}^{\mathrm{act}}$ of the maximum standardized deviation across the $m$ coordinates, which reduces $c_{0.95}^{\mathrm{act}}$ from $3.1816$ ($m_{\mathrm{cal}}=54$) to $2.2351$ ($m=2$ for $\mathrm{do}(X_3=0)$).

For $\mathrm{do}(X_3=0)$ at $v=(1,2,1,2)$, the active outcome probabilities are $\alpha_Y=\pr(Y=2\mid X_1=1,X_2=2,X_3=1)$ and $\beta_Y=\pr(Y=2\mid X_1=1,X_2=2,X_3=0)$, with survey-weighted point estimates $\widehat\alpha_Y=0.681956$ and $\widehat\beta_Y=0.335808$, giving the plug-in sharp interval $[\max(0,1-\widehat\beta_Y/\widehat\alpha_Y),\,\min\{1,(1-\widehat\beta_Y)/\widehat\alpha_Y\}]=[0.5076,0.9740]$.
The clustered survey design can induce covariance between estimates from distinct parent cells. Let $\widehat\Sigma$ denote the $2\times 2$ design covariance submatrix of $(\widehat\alpha_Y,\widehat\beta_Y)^\T$ extracted from $\widehat\Sigma_{\mathrm{surv}}$.
For a candidate ratio $r\geq 0$, the estimated variances of the linear forms $\widehat\beta_Y-r\widehat\alpha_Y$ and $1-\widehat\beta_Y-r\widehat\alpha_Y$ are $s_1^2(r)=(-r,1)\widehat\Sigma(-r,1)^\T$ and $s_2^2(r)=(r,1)\widehat\Sigma(r,1)^\T$.
With $z_{0.975}$ denoting the $97.5$th percentile of the standard normal distribution, the upper confidence limits
\[
 \begin{aligned}
 r_1^U&=\sup\bigl\{r\geq0:\widehat\beta_Y-r\widehat\alpha_Y+z_{0.975}s_1(r)\geq0\bigr\},\\
 r_2^U&=\sup\bigl\{r\geq0:1-\widehat\beta_Y-r\widehat\alpha_Y+z_{0.975}s_2(r)\geq0\bigr\}
 \end{aligned}
\]
(with $r_k^U=\infty$ if the corresponding set is unbounded) each have asymptotic noncoverage probability at most $0.025$, so $[\max(0,1-r_1^U),\min(1,r_2^U)]=[0.2698,1.0000]$ forms an asymptotic $95\%$ confidence interval for the identified interval.

\begin{figure}[!htbp]
\centering
\includegraphics[width=.82\linewidth]{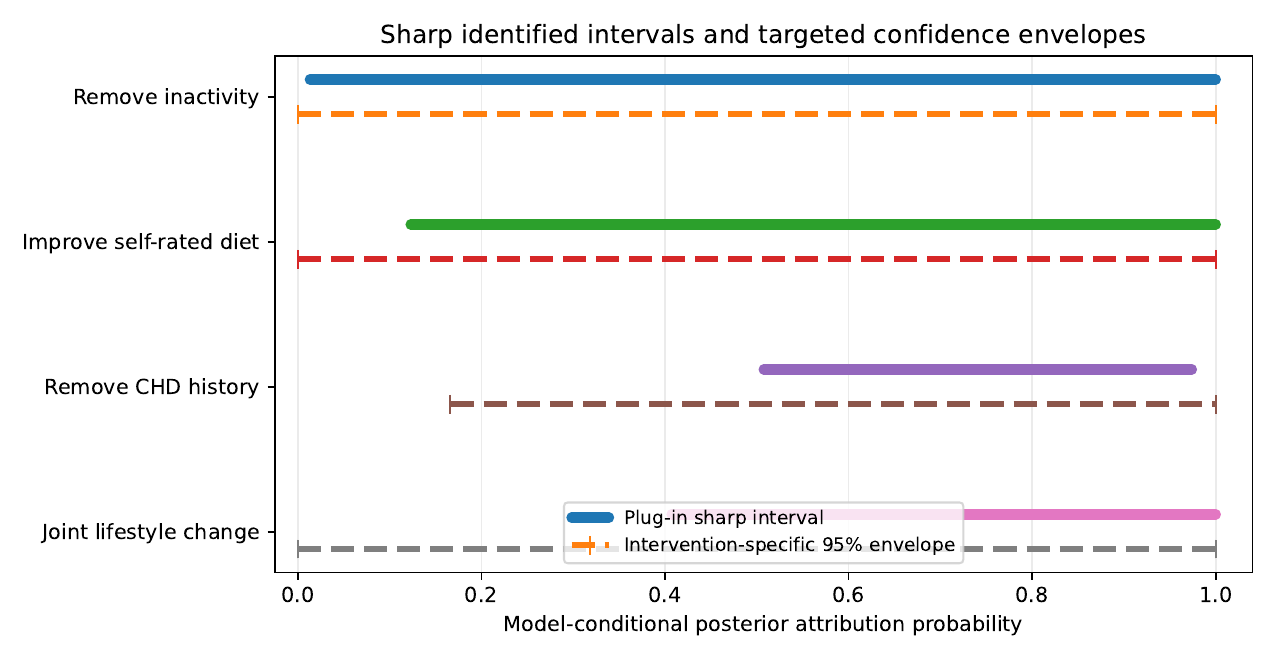}
\caption{Survey-weighted plug-in sharp attribution intervals (solid bars) and active-coordinate 95\% simultaneous Gaussian confidence intervals (dashed bars) for four candidate interventions in NHANES 2017--2018.}
\label{fig:sm-survey-bounds}
\end{figure}

Table~\ref{tab:sm-survey-compat} reports the maximum point violation statistics
\[
\begin{aligned}
\Delta_{\mathrm{mono}}(\widehat F)&=\max_{j,s,k,r}\{\widehat F_j(r;s,x_{k+1})-\widehat F_j(r;s,x_k)\}_+\\
\text{and}\quad\Delta_{\mathrm{inc}}(\widehat F)&=\max_{j,s,k,r}\{\widehat F_j(r-1;s,x_k)-\widehat F_j(r;s,x_{k+1})\}_+
\end{aligned}
\]
for the ordered dietary restrictions in Proposition~\ref{prop:cdf-sandwich}, and Fig.~\ref{fig:sm-survey-bounds} displays the survey-weighted attribution intervals and simultaneous Gaussian confidence intervals.

\begin{table}[!htbp]
\setlength{\tabcolsep}{4pt}
\tbl{Compatibility diagnostics of estimated NHANES margins under ordered dietary response restrictions}
{\begin{tabular}{>{\raggedright\arraybackslash}p{.39\linewidth}*{3}{>{\centering\arraybackslash}p{.17\linewidth}}}
Response restriction & Maximum violation & Point estimates compatible & Confidence-set intersection\\[2pt]
Unrestricted & $0.0000$ & Yes & Yes\\
Monotonicity, $\Delta_{\mathrm{mono}}(\widehat F)$ & $0.1989$ & No & Yes\\
Monotonicity and unit increments, $\max(\Delta_{\mathrm{mono}},\Delta_{\mathrm{inc}})$ & $0.2470$ & No & Yes
\end{tabular}}
\label{tab:sm-survey-compat}
\begin{tabnote}
Maximum violation is the largest positive discrepancy in the distribution-function inequalities \eqref{eq:cdf-sandwich} across mechanisms $X_3$ and $Y$. The final column indicates whether the simultaneous 95\% Gaussian confidence region intersects the compatible margin polytope $\mathcal M$.
\end{tabnote}
\end{table}

\FloatBarrier
\subsection{Additional examples on ordinal exposures and network attribution}
\label{sec:sm-three-level}

Two further examples illustrate bounded adjacent increments for an ordinal exposure and multi-parent attribution in an engineering reliability network.

First, consider a two-variable system $X\to Y$ with ordered exposure $X\in\{x_0<x_1<x_2\}$, outcome $Y\in\{0,1,2\}$, score $g(y)=y$, factual record $(X,Y)=(x_2,2)$, and intervention $\mathrm{do}(X=x_0)$.
Suppose the marginal probabilities $\mu_Y(w;x)=\pr(Y=w\mid X=x)$ for $w=0,1,2$ are $(0.60,0.30,0.10)$ at $x_0$, $(0.40,0.40,0.20)$ at $x_1$, and $(0.20,0.30,0.50)$ at $x_2$, so $\alpha_Y=0.50$.
Writing $q(w)=q_Y(w;x_0)$, Theorem~\ref{thm:localization} gives $q(0)\leq 1.00$, $q(1)\leq 0.60$, and $q(2)\leq 0.20$ without restrictions, so $E(Y_{x_0}\mid V=v)=q(1)+2q(2)\in[0.00,1.00]$ and $\tau_g(x_0;v)\in[1.00,2.00]$.
Under monotonicity $Y_{x_0}\leq Y_{x_1}\leq Y_{x_2}$, $\{Y_{x_0}=2\}\subseteq\{Y_{x_2}=2\}$ fixes $q(2)=0.10/0.50=0.20$, raising the lower mean bound to $0.40$ and narrowing $\tau_g(x_0;v)$ to $[1.00,1.60]$.
Adding the unit-increment restriction $Y_{x_{k+1}}-Y_{x_k}\leq 1$ implies $Y_{x_1}\geq 1$ whenever $Y_{x_2}=2$, so $\pr\{Y_{x_0}=0,Y_{x_2}=2\}\leq\pr\{Y_{x_0}=0,Y_{x_1}\geq 1\}\leq\pr\{Y_{x_0}=0\}-\pr\{Y_{x_1}=0\}=0.20$; dividing by $\alpha_Y=0.50$ gives $q(0)\leq 0.40$ and hence $q(1)\geq 0.40$, further tightening $\tau_g(x_0;v)$ to $[1.00,1.20]$ (Tables~\ref{tab:s-three-level-witness}--\ref{tab:three-level-example}).

\begin{table}[!htbp]
\tbl{Joint response distributions attaining the minimum and maximum conditional outcome means in the three-level example}
{\begin{tabular}{lccc}
Response vector & Monotone minimum & Adjacent-increment minimum & Common maximum\\[2pt]
$(0,0,0)$ & 0.2 & 0.2 & 0.2\\
$(0,0,1)$ & 0   & 0.2 & 0.2\\
$(0,0,2)$ & 0.2 & 0   & 0\\
$(0,1,1)$ & 0   & 0   & 0.1\\
$(0,1,2)$ & 0.1 & 0.2 & 0.1\\
$(0,2,2)$ & 0.1 & 0   & 0\\
$(1,1,1)$ & 0.3 & 0.1 & 0\\
$(1,1,2)$ & 0   & 0.1 & 0.2\\
$(1,2,2)$ & 0   & 0.1 & 0.1\\
$(2,2,2)$ & 0.1 & 0.1 & 0.1
\end{tabular}}
\label{tab:s-three-level-witness}
\end{table}

\begin{table}[!htbp]
\tbl{Sharp intervals in the three-level exposure and outcome example}
{\begin{tabular}{lcc}
Response restriction & $E(Y_{x_0}\mid V=v)$ & $\tau_g(x_0;v)$\\[2pt]
Unrestricted & $[0.00,1.00]$ & $[1.00,2.00]$\\
Monotonicity & $[0.40,1.00]$ & $[1.00,1.60]$\\
Monotonicity and adjacent increments at most one & $[0.80,1.00]$ & $[1.00,1.20]$
\end{tabular}}
\label{tab:three-level-example}
\end{table}

\begin{figure}[!htbp]
\centering
\includegraphics[width=.95\linewidth]{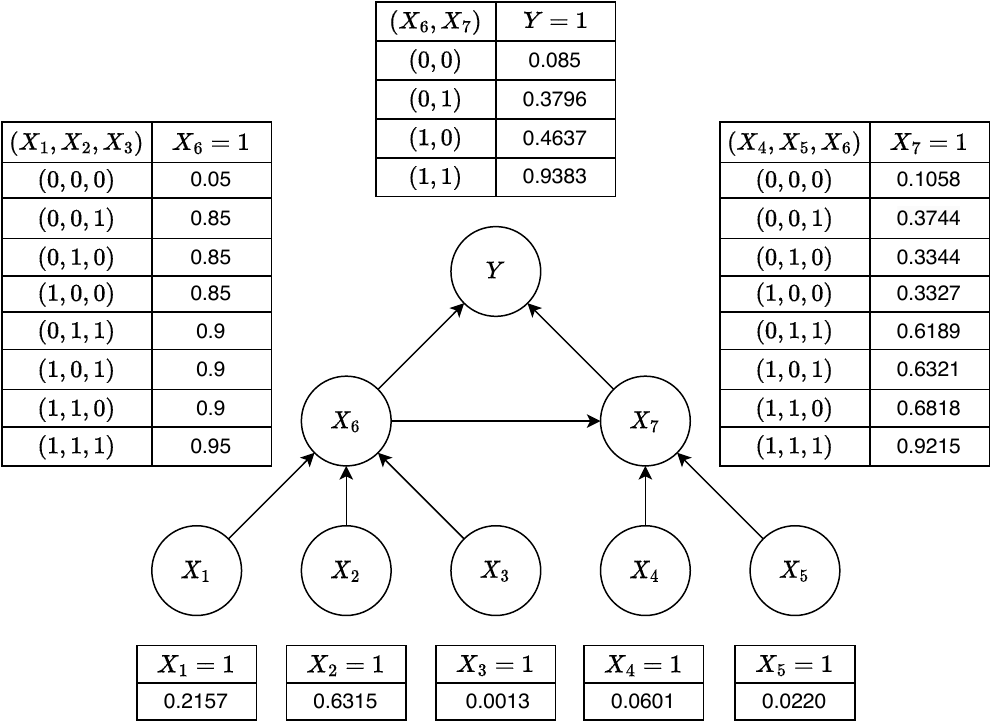}
\caption{Causal reliability network and conditional failure probabilities for the harmonic gear drive, reproduced from \citet{Yang2021}.}
\label{fig:s-hgd-physical}
\end{figure}

Second, consider the harmonic gear drive reliability network of \citet{Yang2021} shown in Fig.~\ref{fig:s-hgd-physical}, with binary component failure indicators $X_1,\ldots,X_5$, subsystem failure indicators $X_6,X_7$, system failure indicator $Y$, and parent sets $\Pa(X_6)=\{X_1,X_2,X_3\}$, $\Pa(X_7)=\{X_4,X_5,X_6\}$, and $\Pa(Y)=\{X_6,X_7\}$.
Conditional on all variables failing ($V=\boldsymbol1$), the factual probabilities are $\alpha_{X_6}=0.95$, $\alpha_{X_7}=0.9215$, and $\alpha_Y=0.9383$.
Table~\ref{tab:s-hgd} reports the unrestricted sharp attribution intervals for individual and joint component repairs alongside the point-identified effects under structural monotonicity, which coincide with the unrestricted lower bounds by Corollary~\ref{cor:monotone-endpoint}.

\begin{table}[!htbp]
\tbl{Unrestricted attribution intervals and point-identified effects under structural monotonicity in the reliability network}
{\begin{tabular}{lcc}
Component or subsystem repair & Unrestricted sharp interval & Effect under monotonicity\\[2pt]
$X_1$, direct-current motor & $[0.0356,0.0810]$ & $0.0356$\\
$X_2$, wave generator & $[0.0356,0.0810]$ & $0.0356$\\
$X_3$, convex gear & $[0.0356,0.0810]$ & $0.0356$\\
$X_4$, flexspline & $[0.1661,0.2364]$ & $0.1661$\\
$X_5$, circular spline & $[0.1589,0.2282]$ & $0.1589$\\
$X_6$, wave-generation subsystem & $[0.6771,0.7696]$ & $0.6771$\\
$X_7$, gear-transmission subsystem & $[0.5058,0.5716]$ & $0.5058$\\
$(X_4,X_5)$, joint repair & $[0.3003,0.3880]$ & $0.3003$\\
$(X_1,X_2,X_3)$, joint repair & $[0.6415,0.7696]$ & $0.6415$
\end{tabular}}
\label{tab:s-hgd}
\end{table}

\FloatBarrier

\end{document}